\documentclass[oneside]{article}

\usepackage[T1]{fontenc} 
\usepackage[english]{babel} 
\usepackage[hmarginratio=1:1,top=32mm,columnsep=20pt]{geometry} 
\usepackage{graphicx}
\usepackage{fancyhdr} 

\usepackage{titling} 
\usepackage{nth}
\usepackage{nameref, hyperref} 

\usepackage{amsmath,amssymb}

\usepackage{changepage}

\usepackage[utf8x]{inputenc}

\usepackage[right]{lineno}

\usepackage{array}
\newcolumntype{H}{>{\setbox0=\hbox\bgroup}c<{\egroup}@{}}

\usepackage{booktabs} 

\usepackage{tabularx}
\usepackage{multirow} 
\usepackage{pdflscape} 
\usepackage{longtable} 

\usepackage{ogonek} 

\usepackage{float} 

\usepackage{multicol,subcaption}

\usepackage{tikz}

\usepackage{nameref}

\usepackage{chngcntr}
\usepackage{soul}

\usepackage{verbatim}

 \usepackage[round, authoryear]{natbib}
 
\usepackage{longtable}
\usepackage{tabu} 

\definecolor{cc686e9}{RGB}{198,134,233} 
\definecolor{c5fc613}{RGB}{95,198,19} 
\definecolor{c00caff}{RGB}{0,202,255} 
\definecolor{cea8615}{RGB}{234,134,21} 
\definecolor{c2e9072}{RGB}{46,144,114} 
\definecolor{cff5c81}{RGB}{255,92,129} 

\definecolor{cdf89ff}{RGB}{223,137,255} 
\definecolor{c73c000}{RGB}{115,192,0} 
\definecolor{c00c4ff}{RGB}{0,196,255} 
\definecolor{c4c463e}{RGB}{76,70,62} 
\definecolor{cff8805}{RGB}{255,136,5} 
\definecolor{cff5584}{RGB}{255,85,132} 
\definecolor{c00bd94}{RGB}{0,189,148} 
\definecolor{cd3b3b0}{RGB}{211,179,176} 

\definecolor{cf2ff00}{RGB}{242,255,0} 
\definecolor{c7e00c2}{RGB}{126,0,194} 
\definecolor{cc0c0c0}{RGB}{192,192,192} 

\definecolor{c7495cb}{RGB}{116,149,203} 
\definecolor{cffbcbc}{RGB}{255,188,188} 
\definecolor{c8a8a8a}{RGB}{138,138,138} 

\definecolor{cda027f}{RGB}{218,2,127} 
\definecolor{c29033c}{RGB}{41,3,60} 
\definecolor{c639a76}{RGB}{99,154,118} 
\definecolor{c36f230}{RGB}{54,242,48} 
\definecolor{cf9f99d}{RGB}{249,249,157} 
\definecolor{cbae396}{RGB}{125,232,140} 
\definecolor{c9ee0e0}{RGB}{158,224,224} 
\definecolor{ca5bdcc}{RGB}{165,189,204} 
\definecolor{c4cdc38}{RGB}{76,220,56} 
\definecolor{cd340bd}{RGB}{211,64,189} 
\definecolor{c874010}{RGB}{135,64,16} 
\definecolor{c666699}{RGB}{102, 102, 153} 

\definecolor{c4b771a}{RGB}{75,119,26} 
\definecolor{cda0c5a}{RGB}{218,12,90} 
\definecolor{c7e3f9c}{RGB}{126,63,156} 
\definecolor{c501653}{RGB}{80,22,83} 
\definecolor{c7de88c}{RGB}{125,232,140} 
\definecolor{ce1ac06}{RGB}{225,172,6} 
\definecolor{c430337}{RGB}{67,3,55} 
\definecolor{c6e4b83}{RGB}{110,75,131} 
\definecolor{ce40030}{RGB}{228,0,48} 
\definecolor{c102ee2}{RGB}{16,46,226} 

\definecolor{c23966F}{RGB}{35,150,111} 
\definecolor{c48D16C}{RGB}{72,209,108} 
\definecolor{cFF7045}{RGB}{255,112,69} 
\definecolor{c00C7FF}{RGB}{0,199,255} 

\newcolumntype{+}{!{\vrule width 2pt}}

\newlength\savedwidth

\newcommand\thickhline{\noalign{\global\savedwidth\arrayrulewidth\global\arrayrulewidth 2pt}%
\hline
\noalign{\global\arrayrulewidth\savedwidth}}

\usepackage[aboveskip=1pt,labelfont=bf,labelsep=period,justification=raggedright,singlelinecheck=off]{caption}

\title{Clusters of investors around Initial Public Offering} 

\author{
 \textsc{Margarita Baltakien{\.e}}\thanks{Corresponding author}\\[1ex]
 \normalsize Tampere University \\ 
 \normalsize \href{mailto:margarita.baltakiene@tuni.fi}{margarita.baltakiene@tuni.fi} 
 \and 
 \textsc{K\k{e}stutis Baltakys} \\
 \normalsize Tampere University \\ 
 \and 
 \textsc{Juho Kanniainen}\\ 
 \normalsize Tampere University \\ 
 \and  
 \textsc{Dino Pedreschi}\\
 \normalsize University of Pisa \\ 
 \and  
 \textsc{Fabrizio Lillo}\\
 \normalsize University of Bologna \\ 
}

\begin{document}

\noindent
\maketitle
\footnotemark{This is a pre-print of an article published in Palgrave Communications.
The final authenticated version is available online at: \href{https://doi.org/10.1057/s41599-019-0342-6}{https://doi.org/10.1057/s41599-019-0342-6} }

\abstract
The complex networks approach has been gaining popularity in analysing investor behaviour and stock markets, but within this approach, initial public offerings (IPO) have barely been explored. 
We fill this gap in the literature by analysing investor clusters in the first two years after the IPO filing in the Helsinki Stock Exchange by using a statistically validated network method to infer investor links based on the co-occurrences of investors’ trade timing for 69 IPO stocks.
Our findings show that a rather large part of statistically similar network structures form in different securities and persist in time for mature and IPO companies.
We also find evidence of institutional herding.

\section*{Introduction}
Initial public offerings (IPOs) play an important role in financial markets because they open new investment opportunities, redistribute funds’ allocations and attract new investors to the market.
An IPO is usually a long-awaited event in the life of a privately held company, both for the current stockholders and the public exchange investors, giving the owners the opportunity to cash in and giving the investors a chance to gain from potential underpricing and future returns.
Here, numerous financial studies have addressed various behavioural biases in relation to IPOs: 
\citet{ljungqvist2005does} analysed the satisfaction with an IPO underwriter's performance, \citet{ljungqvist2003ipo} indicated a unique pricing behaviour around the dot-com bubble, while \citet{kaustia2008investors} found that investors’ personal experiences and previous IPO returns have a significant impact on future IPO subscriptions. Other studies have analysed IPO investments {\citep{karhunen2001shareownership}}, IPO earnings \citep{spohr2004earnings} and IPO underpricing \citep{keloharju1993winner} in financial markets on an aggregated level. 

Financial markets, in turn, are complex systems comprised of financial decisions, information flows and direct and indirect investor interactions.
A typical aspect of a financial market is multidimensionality and agent heterogeneity \citep{lakonishok1990weekend, musciotto2016patterns}.
Making an investment decision is a complex procedure because it is layered with different choices that are influenced by various market factors, investors’ experiences, wealth and investors’ stage of life.
It is crucial to understand the characteristics of the underlying investor behaviour patterns because these, when combined with their behaviours, shape the dynamics of the whole market and thus are important factors in explaining the booms and bubbles in the financial markets \citep{ranganathan2018dynamics}.
Because investors seek higher returns, one possibility is to use social networks and other private information channels to follow other investors’ strategies and to exploit privately channelled information in stock markets.  
Recently, \citet{baltakys2018neighbors} provided evidence of the negative relationship between distance and trade timing similarity for household investors, indicating that face-to-face communication is still important in financial decision making. 
According to \citet{ozsoylev2013investor}, information links can be identified from realised trades because investors who are directly linked in the information network tend to time their transactions similarly.  
We follow this idea and use observations on investor-level transactions from shareholder registration data to identify the links between investors, here with a special focus on identifying investor clusters. 
Prior studies have investigated the structures of investor networks in different contexts \citep{ozsoylev2013investor,tumminello2012identification,gualdi2016statistically,musciotto2018long,ranganathan2018dynamics,baltakys2018multilayer}, but investor clusters around IPOs have barely been explored.

We address this research gap by performing a broad multistock exploratory analysis of investor clusters over 69 stocks in the first two years of their IPO. 
In particular, we seek to establish whether the identified investor clusters are persistent over the first two years of the IPOs and appear across multiple IPO securities, as well as with existing, mature stocks in the market.
Our analysis unveils statistically robust investor clusters that form simultaneously in various securities, and that persist over time.

Most of the earlier papers perform analyses on an aggregated category level \citep{karhunen2001shareownership,grinblatt2001distance,lillo2015news,siikanen2018facebook} or concentrate on a single highly liquid stock \citep{tumminello2012identification, musciotto2018long}. 
Even though earlier studies might have included nearly all market participants \citep{tumminello2011statistically, musciotto2018long}, due to the focus on a single most liquid security, the results were limited and insufficient to conclude what strategies investors employ when trading over multiple securities.
In contrast to previous research in the IPO literature, the current study is the first one on early-stage trading behaviour patterns on an individual investor account level.
On the other hand, in opposition to the existing research on investor networks, in the current paper, instead of focusing of heavily capitalised stocks we analyse collective investor trading strategies that emerge after IPOs in the Helsinki Stock Exchange (HSE).

With the growing amounts of data and the availability of new datasets, the network theory has become a popular approach in analysing financial complex systems (e.g., \citep{emmert2017computational}). 
Notwithstanding the high interest in the market structure, investor networks and the complexity of investor behavioural interrelationships remain weakly explored.
Indeed, high precision financial investor-level datasets covering years of historical data and containing information about the social links are very rare and expensive because of their sensitive nature. 
Moreover, transactional data often have no explicit or implicit links between investors.
As a consequence, the network inference methodologies have gained much interest in recent research \citep{ozsoylev2013investor, gualdi2016statistically}.
Similar to \citet{musciotto2018long}, we use the statistical validation method proposed by \citet{tumminello2011statistically}, which best suits our objectives and the available dataset.

In the current paper, we infer investor networks based on the investors’ trading co-occurrences for 69 securities that had their IPOs between the years 1995 and 2007, and we obtain multilink networks covering two years after their IPOs. 
Further, by applying the Infomap algorithm \citep{rosvall2008maps} on the investor networks, we obtain clusters of investors that share high trade-timing synchronisation.
With the obtained network partitioned into clusters, we detect statistically robust clusters that persist in the networks between the first and the second years after the IPO.
We also find clusters that form and re-occur over multiple securities. 
Finally, by cross-validating investor clusters on IPO securities with the investor clusters of more mature stocks, we conclude that the phenomenon of persistent clusters observed in earlier studies (see e.g. \citep{musciotto2018long}) is not limited to mature companies but is also observable in young securities during the first years after their IPO.

\section*{Dataset and methodology} 
\subsection*{Dataset} \label{sec:dataset}
In this paper, we use a unique database provided by Euroclear Finland.   
The dataset contains all transactions executed in the Helsinki Stock Exchange by Finnish stocks shareholders between 1995 and 2009 on a daily basis.
The data records represent the official certificates of ownership and include all the transactions executed in the Helsinki Stock Exchange that change an ownership of assets.
Each transaction in the dataset has a rich set of attributes – such as investor sector code, investor birth year, gender and postal code – that we make use of in our analysis to identify and characterise the investor groups.
The dataset classifies investors into six main categories: households; nonfinancial corporations; financial and insurance corporations; government; nonprofit institutions; and the rest of the world. 
Finnish domestic investors correspond to a separate account ID, while foreign investors can choose the nominee registration for the trades.  
However, the analysis cannot be conducted for nominee-registered transactions because individual nominee investors cannot be uniquely identified. 
Rather, the nominee investors are pooled together under the custodian’s nominee trading account. 
Therefore, a single nominee-registered investor’s account holdings may correspond to a large aggregated ownership of several foreign investors. 
So to avoid inconsistencies in the results, we eliminated nominee transactions from our analysis.
This dataset has been also analysed and described in previous research (e.g., \citep{ilmanen1999shareownership,baltakys2018multilayer,ranganathan2018dynamics,baltakys2018neighbors,siikanen2018facebook}).

The analysed data are restricted to marketplace transactions for securities that had their IPO listing in the Helsinki Stock Exchange between 1995 and 2009.  
The official listing dates were provided by NASDAQ OMX Nordic explicitly for the current research.  
We analyse 69\footnote{In total, 75 securities had their IPOs during our analysis period. 
In this study we estimate investor networks during a two-year period after their IPO date; therefore, we discarded ISIN FI0009014716 because its two-year period falls out of our analysis period. 
Additionally, five ISINs (FI0009015580, FI0009015291, FI0009015713, FI0009005250 and FI0009902514) were discarded from the analysis, because no networks were estimated for them.}\footnote{
Unfortunately, the data appear to have issues with the trading date attribute for some securities, particularly for the transactions between 1998 and 2004.
The net trading volumes on a daily resolution do not reconcile to 0 for all trading dates, while the volume sold should be equal to the volume bought per each stock during each day across all investors. 
This suggests that some transactions in the dataset were misplaced timewise because of incorrectly recorded trading dates.
Only 14 of 69 securities fall into the completely error-free data period, and are marked in bold in Table \ref{tab:num_trans}.
} stocks in total that were listed in Finland on the Main Exchange or First North in the given time period (Table \ref{tab:num_trans}).
Some companies (e.g. Oriola) have two share classes with different voting rights. 
Class A shares give the owner more voting rights than Class B and hence potentially falls under a separate group of investors. 
Therefore, the comparison or a direct substitution of shares with one another seems improper, and we consider the securities with different voting classes as separate stocks.

\begin{table}[H]
\caption{
{\bf Summary of IPO stocks.}  International Securities Identification Number (ISIN), company, industry, total number of transactions, total number of unique investors and the IPO day of the security.
ISINs from the error-free set are marked in bold.}
\centering
\resizebox{!}{0.46\textheight}{
\begin{tabular}{lllrrl}
\hline
  ISIN & company name & industry & \begin{tabular}{r}total \# of\\  transactions\end{tabular}   & \begin{tabular}{r}\# of unique \\  investors\end{tabular} & IPO date \\
\thickhline
\textbf{FI0009004881} & Aspoyhtym{\"a} & Industrials & 13157 & 2070 & 1995-01-12 \\
FI0009800346 & Orion B & Basic Materials & 399268 & 45588 & 1995-05-11 \\
FI0009800320 & Orion A & Basic Materials & 116334 & 18132 &1995-05-11 \\
FI0009900336 &Lemmink{\"a}inen & Industrials &94849 & 13269 &1995-06-01 \\
FI0009005318 &Nokian Renkaat & Consumer Goods & 1152852 & 60476 &1995-06-01 \\
FI0009800643 &YIT & Industrials & 896718 & 54808 &1995-09-04 \\
FI0009005870 &Konecranes & Industrials & 715306 & 26940 &1996-03-27 \\
FI0009005953 &Stora Enso A & Basic Materials &73993 & 14816 &1996-05-02 \\
FI0009005961 &Stora Enso R & Basic Materials &1514604 & 52567 &1996-05-02 \\
FI0009005987 & UPM-Kymmene & Basic Materials &2323897 &118769 &1996-05-02 \\
FI0009006381 & PKC Group & Industrials & 194480 & 24624 &1997-04-03 \\
FI0009006415 &Nordic Aluminium & Basic Materials &19012 &4291 &1997-04-24 \\
FI0009005805 & Kyro & Consumer Services &44418 &9100 &1997-06-09 \\
FI0009006589 &Rocla & Basic Materials &15415 &3918 &1997-06-17 \\
FI0009006621 & Helsingin Puhelin & Telecommunications & 116532 & 32367 &1997-11-25 \\
FI0009006738 &Elcoteq & Technology & 503265 & 43323 &1997-11-26 \\
FI0009006696 & P{\"o}yry & Industrials & 125202 & 14135 &1997-12-02 \\
FI0009006761 & Mets{\"a} Tissue & Basic Materials &11286 &3725 &1997-12-09 \\
FI0009007017 &Alma Media I & Consumer Services &10673 &2472 &1998-04-01 \\
FI0009007025 & Alma Media II & Consumer Services &30500 &5383 &1998-04-01 \\
FI0009007066 &Ramirent & Industrials & 295726 & 21662 &1998-04-30 \\
FI0009006829 & Sponda & Financials & 213977 & 19500 &1998-06-01 \\
FI0009007215 & Mandatum Pankki & Financials &25732 &6430 &1998-08-03 \\
FI0009007264 &Elektrobit & Technology & 681542 & 74839 &1998-09-15 \\
FI0009007371 & Sonera & Telecommunications &1504103 &140253 &1998-11-17 \\
FI0009007355 &Rapala VMC & Consumer Goods &30739 &5202 &1998-12-04 \\
FI0009007132 & Fortum & Utilities &2068556 &120902 &1998-12-18 \\
FI0009007629 & Conventum & Financials &13395 &2736 &1999-03-01 \\
FI0009801286 & Janton & Consumer Services &22946 &5418 &1999-03-15 \\
FI0009007553 & Eimo & Telecommunications & 187912 & 24664 &1999-03-23 \\
FI0009007728 &Teleste & Technology & 209132 & 22537 &1999-04-06 \\
FI0009007546 &Keskisuomalainen & Consumer Services &11019 &2046 &1999-04-19 \\
FI0009007686 &SanomaWSOY A & Consumer Services &10784 &2438 &1999-05-03 \\
FI0009007694 & Sanoma & Consumer Services & 458541 & 33242 &1999-05-03 \\
FI0009006886 & Technopolis & Financials & 85510 &8892 &1999-06-08 \\
FI0009007819 & Perlos & Telecommunications & 520835 & 44281 &1999-06-28 \\
FI0009007835 &Metso & Industrials &1528914 & 69361 &1999-07-01 \\
FI0009007884 &Elisa & Telecommunications &1209330 &199530 &1999-07-01 \\
FI0009008080 &Aspocomp Group & Industrials &99023 & 10948 &1999-10-01 \\
FI0009007918 & Aldata Solution & Technology & 253021 & 22840 &1999-10-27 \\
FI0009801310 &F-Secure & Technology & 578978 & 70994 &1999-11-09 \\
FI0009008221 &Comptel & Telecommunications & 529255 & 65050 &1999-12-13 \\
FI0009902530 & Nordea Bank & Financials &1081900 &149790 &2000-01-31 \\
FI0009008924 & Sievi Capital & Financials & 91541 & 12109 &2000-05-24 \\
FI0009008833 &Tekla & Telecommunications &73328 &8581 &2000-05-24 \\
FI0009009146 &Tecnomen & Telecommunications &19745 &4532 &2000-07-04 \\
FI0009009054 &Okmetic & Telecommunications &75944 & 10430 &2000-07-05 \\
FI0009009633 & Evox Rifa Group & Telecommunications &51493 & 10203 &2000-11-01 \\
FI0009009567 &Vacon & Telecommunications& 80081 & 10770 &2000-12-19 \\
FI0009008270 &SSH Comm. Security & Technology & 112633 & 16433 &2000-12-22 \\
FI0009009674 & AvestaPolarit & Basic Materials &24752 &4299 &2001-01-30 \\
FI0009009377 & CapMan & Financials & 74153 & 11279 &2001-04-02 \\
FI0009010219 &Glaston & Industrials & 47748 &8174 &2001-04-02 \\
FI0009010854 &Lassila \& Tikanoja & Industrials & 120822 & 13385 &2001-10-01 \\
FI0009010862 &Suominen & Consumer Goods &51734 &7052 &2001-10-01 \\
SE0000667925 & Telia & Telecommunications & 870709 &107088 &2002-12-09 \\
\textbf{SE0000110165} &OMX &Financials& 8721 &1851 &2003-09-04 \\
\textbf{FI0009012843} &Kemira GrowHow &Basic Materials & 142417 & 25253 &2004-10-18 \\
\textbf{FI0009013296} & Neste Oil &Oil \& Gas&1387293 & 81750 &2005-04-21 \\
\textbf{FI0009013429} &Cargotec &Industrials & 474949 & 29210 &2005-06-01 \\
\textbf{FI0009013312} &Affecto &Technology&40635 &5726 &2005-06-01 \\
\textbf{FI0009013403} & Kone &Industrials & 618717 & 30192 &2005-06-01 \\
\textbf{FI0009013924} &Salcomp &Industrials &28721 &3688 &2006-03-17 \\
\textbf{FI0009010391} &Ahlstrom &Basic Materials &87853 & 16594 &2006-03-17 \\
\textbf{FI0009013593} & FIM Group &Financials&11379 &3084 &2006-04-21 \\
\textbf{FI0009014344} &Oriola A &Health Care &25922 &5595 &2006-07-03 \\
\textbf{FI0009014351} &Oriola B &Health Care & 116890 & 19279 &2006-07-03 \\
\textbf{FI0009012413} &Terveystalo Health &Health Care &35203 &8946 &2007-04-10 \\
\textbf{FI0009015309} &SRV Yhti{\"o}t &Industrials &56384 &9579 &2007-06-15 \\
\hline
\end{tabular}}
\label{tab:num_trans}
\end{table}

Table \ref{tab:summary} gives the number of investors, the number of transactions and the traded volume for the entire set of 69 IPO stocks. 
The total number of investors who traded an IPO security is 570,039, and the total number of transactions is 76,505,089.
The table also shows the number of nominee and non-nominee-registered investors.
As shown, a few nominee accounts perform roughly twice as many trades as the non-nominee accounts.
\begin{table}[!ht]
\centering
\caption{
{\bf Summary of the number of investors, absolute exchanged shares volume and the number of transactions.} Note that the total volume in the table is counted twice, both for the selling and buying transactions. 
Here, 43 out of 89 investors with a nominee-registered holding type also made transactions with a non-nominee-registered holding type.}
\resizebox{1\textwidth}{!}{
\begin{tabular}{lrrrHHH}
\hline

                    Investor category &	\# ids & volume & \# transactions\\ 
\thickhline
Non-financial corporations &   29,008 &  10,492,715,279    & 3,678,419  & 8,363
 &  1,049,640,371  &  493,666 \\ 
Financial and insurance corporations &  	 827 &  350,594,504,886  & 55,735,780 & 750
 &  23,307,610,030 & 8,149,556  \\    
Government &  	 277 &  7,279,324,503    & 298,434   & 113
 &  590,306,199   &  31,099 \\ 
Households &  532,387 &  8,984,345,323    & 12,965,717  & 121,434
 & 735,428,351   & 1,555,604 \\ 
Non-profit institutions &    3,407 &  937,609,174     & 291,922   & 1,067
 &  78,654,895    &  38,824\\ 
Rest of the world &    4,133 &  12,505,262,104    & 3,534,817  & 806 
 &  1,012,077,566 &  563,797 \\ 
\textbf{Total} & \textbf{570,039}  &  \textbf{390,793,761,269}      & \textbf{76,505,089} & \textbf{132,533} & \textbf{26,773,717,412} & \textbf{10,832,546} \\ 
Nominee registered &  89 	 &  331,154,383,799 & 51,782,691 & 71 & 21,987,309,532 & 7,566,437  \\ 
Non-nominee registered &  569,993 &  59,639,377,470  & 24,722,398 & 132,495  & 4,786,407,880 & 3,266,109  \\  
   \hline
\end{tabular}}
\label{tab:summary}
\end{table}

\subsubsection*{Methodology}
The given dataset is composed of transaction data where investors’ social links are not explicitly given, nor can they be directly obtained from other sources because of data anonymisation. 
However, given that investors must individually react and adapt to a quickly changing environment, they should identify and follow the best trading strategies.
To detect investors with {\em similar} trading strategies or, more precisely, trade timing similarity, we take a look at the pairwise investors’ trading co-occurrences.  
In the current paper, we use a statistically validated network (SVN) method first introduced by \citet{tumminello2011statistically}. 
This method, briefly presented below, has been demonstrated to be effective in investigating financial, biological and social systems \citep{tumminello2011statistically,tumminello2012identification}. 

To compare the trading position taken by an investor on a given day, irrespective of the absolute volume traded, a categorical variable is introduced that describes the investor’s trading activity. 
For each investor $i$ and each trading day $t$ having the volume sold of a security $V_s(i,t)$ and the volume bought of a security $V_b(i,t)$, we calculate the scaled net volume ratio as follows:
\begin{eqnarray}
r(i,t)=\frac{V_b(i,t)-V_s(i,t)}{V_b(i,t)+V_s(i,t)}
\label{eq:cat_ratio}
\end{eqnarray} 
Then, a daily trading state can be assigned for an investor after having selected a threshold $\theta$, as follows:
\begin{eqnarray*}
\begin{cases}
  b - \text{primarily buying state, when }  r(i,t)>\theta      \\
  s - \text{primarily selling state, when }	r(i,t)<	{-\theta} \\ 
  bs - \text{buying and selling state, when } {-\theta} \leq r(i,t) \leq \theta
\end{cases}
\end{eqnarray*}
Note that $r(i,t)$ is not defined for day $t$ that had no trading activity, and therefore, no trading state is assigned.
In our analysis, much like in \citet{musciotto2016patterns}, we set $\theta = 0.25$. 
We have verified that the calculations are not sensitive to $\theta$ selection: the results do not vary significantly for the $\theta$ threshold ranging from 0.01 to 0.25. 
With this categorisation, the system can be mapped into a bipartite network. 
We will take one set of nodes composed of investors and the other set composed of the trading days.

The states $b$, $s$ and $bs$ of investor $i$ are indicated as $i_b$, $i_s$ and $i_{bs}$, respectively. 
There are nine possible combinations of the three trading states between investors $i$ and $j$: ($i_b$,$j_b$), ($i_b$,$j_s$), ($i_b$,$j_{bs}$), ($i_s$,$j_b$), ($i_s$,$j_s$), ($i_s$,$j_{bs}$), ($i_{bs}$,$j_b$), ($i_{bs}$,$j_s$) and ($i_{bs}$,$j_{bs}$).
Because we are focusing on the positive relationship between investors’ trading strategies, we  further analyse only the situations where both investors have been in a buy state ($i_b$,$j_b$), both investors have been in the sell state ($i_s$,$j_s$), and both investors have been day traders ($i_{bs}$,$j_{bs}$), thus excluding the other six trading state co-occurrences.

\subsubsection*{Statistically validated networks}
With the categorical variables on the trading states, the co-occurrence of the trading states of investors $i$ and $j$ can be identified and statistically validated. 
First, for each investor, her or his activity period is identified. 
Second, for an investor pair, the length of a joint trading period is determined, $T$, which is equal to the number of trading days in an annual data sample for a given security ($\approx$ 250).  
Then, in the intersection periods of a trader’s activity, $N_i^P$ ($N_j^P$) denotes the number of days when investor $i$ ($j$) is in a given state $\{b, s, bs\}$. 
Moreover, $N_{i,j}^P$ denotes the number of days when we observe the co-occurrence of the given states for investors $i$ and $j$.
Under the null hypothesis of the random co-occurrences of a state for investors $i$ and $j$, the probability of observing $X$ co-occurrences of the investigated states for two investors in $T$ observations can be expressed by the hypergeometric distribution $H(X|T, N_i^P, N_j^P)$ \citep{tumminello2011statistically}. For each trading state $P = \{b, s, bs\}$, a $p$-value can be associated as follows:
\begin{eqnarray}
p\left(N_{i,j}^P\right) = 1 - \sum_{X=0}^{N_{i,j}^P-1}H(X|T, N_i^P, N_j^P)
\label{eq:hyper_pvalue}
\end{eqnarray}

Using the SVN method, for each security we construct two subsequent year networks.  
The analysis for each security spans from the initial listing day up to the second year after the IPO. 
We assign the categorical variables that define the investor’s daily trading state, and we select only domestic Finnish investors who have traded an IPO stock at least five days during the first or second year.
For each analysed security, we take two consecutive one-year periods of categorised trading states for investors.
Taking the projection of the investor set in a year, we obtain an annual monopartite investor network, and two investor networks for consecutive years are obtained for each security.

We adjust the $p$-thresholds using a false discovery rate (FDR) correction \citep{benjamini1995controlling} by taking the sorted $p$-values $p_1 < p_2 < \ldots< p_{n_{tests}} $ in an increasing order and retain those that satisfy $p_i < \alpha \cdot i/n_{\text{tests}}$, $i = 1, \ldots, n_{\text{tests}}$. 
Here, we apply $\alpha$ = 0.05, and $n_{\text{tests}}$ equals the total number of observed relationships in a year.
All networks are essentially multilink networks, where each link describes the type of trading co-occurrence between an investor pair. 
This adjustment is needed because there are multiple links and thus multiple tests with a given network. 
The link between investors $i$ and $j$ is considered to be statistically significant and thus existing if the corresponding $p$-value, $\left(N_{i,j}^P\right)$,  is below the FDR-adjusted $p$-threshold. 
In this way, we obtain validated networks for the first and second years.
As an example, Fig. C.1 in Appendix C shows the first year sorted p-values and the FDR thresholds for Kemira GrowHow links.

\subsubsection*{Statistically validated clusters: persistence in time}
We are interested in the investors’ cluster evolution over time. 
In other words, we want to verify whether investors systematically synchronise their trading strategies with other investors and if such behaviour can be detected in the subsequent year networks. 
With the community partition for each network, we identify persistent clusters (i.e., clusters that share the same statistically significant component of investors in both the first and the second years after the IPO). 
Further, we briefly present the method from \citet{marotta2015bank}.

We are interested in identifying statistically similar clusters that emerged in both years (i.e., clusters with the overexpression of the same investor composition in both clusters, which share nonrandom elements).  
The probability that $X$ elements in the cluster $C_1$ of the first year network composed of $N_{C_1}$ elements also appear in the cluster $C_2$ of the second year composed of $N_{C_2}$ elements under the null hypothesis that the elements in each cluster are randomly selected is given by the hypergeomteric distribution $H(X|N, N_{C_1}, N_{C_2})$, where $N$ is the total number of unique elements over two years. 
By using this distribution, a $p$-value can be associated with the observed number $N_{{C_1}{C_2}}$ of elements of the cluster ${C_1}$ reoccurring in ${C_2}$ according to the following equation:
\begin{eqnarray}
p(N_{{C_1}{C_2}}) = 1 - \sum_{X=0}^{N_{{C_1}{C_2}}-1}H(X|N, N_{C_1}, N_{C_2})
\label{eq:hyper_pvalue3}
\end{eqnarray}

We reject the null hypothesis if $p(N_{{C_1}{C_2}})$ is smaller than a given adjusted threshold, in which case we say that the cluster $C_1$ is statistically similar with the cluster $C_2$. 
We adjust the statistical threshold using the FDR correction with $\alpha = 0.05$ and the number of tests being equal to the total number of cluster pairs over two years that shared at least one common element.

\subsubsection*{Statistically validated clusters: similarity across securities}
Additionally, to check if the same cluster exists over multiple securities, we expand the analysis and further look for statistically significant overlapping clusters across all investigated securities. 
Because the IPO event is the alignment point in our analysis, we look for the overlapping clusters in the set of first-year networks and the set of second-year networks separately. 
We again use the method (Eq. \ref{eq:hyper_pvalue3}) for the cluster overlaps to detect clusters with nonrandomly overlapping elements (investors).  
To calculate the $p$-values, we take $N$ equal to the total number of unique investors across all investigated securities in the same year, where $N_{C_1}$ is the number of investors in the cluster ${C_1}$, $N_{C_2}$ is the number of investors in the cluster ${C_2}$, and $N_{{C_1}{C_2}}$ is the number of common investors in both ${C_1}$ and ${C_2}$. 
Again, we adjust the statistical threshold using the FDR correction, where $\alpha = 0.05$ and the number of tests is equal to the total number of cluster pairs within the same year that shared at least one common element.

\subsubsection*{Over- and underexpression of the characterising investor attributes}
To describe the investor clusters from the perspective of the attributes, such as postal code, age, gender or the type of organisation, we again use the hypergeometric test for identifying nonrandom overlap \citep{tumminello2011community}.
Once we obtain a system of $N$ elements partitioned into clusters (communities), we want to characterise each cluster $C$ of $N_C$ elements.  
Each element of the system has a certain number of attributes from a specific class. 
Here, we want to see if the number of elements in the cluster with a specific attribute value is significantly larger than randomly selecting the elements from the total system elements. 
For each attribute $Q$ of the system, we test if $Q$ is over-expressed in the cluster $C$. 
The probability that $X$ elements in cluster $C$ have the attribute $Q$ under the null hypothesis that the elements in the cluster are randomly selected is given by the hypergeomteric distribution $H(X|N, N_C, N_Q)$, where $N_Q$ is the total number of elements in the system with attribute $Q$. 
By using this distribution, a $p$-value can be associated with the observed number $N_{C,Q}$ of elements in cluster $C$ that have the attribute $Q$ analogously with Eq. \ref{eq:hyper_pvalue3}. 
We reject the null hypothesis if the p-value is smaller than a given FRD-adjusted p-threshold, and we then say that the attribute $Q$ is overexpressed in cluster $C$. 
In the FDR-adjustment, the number of tests is equal to the total number of unique attribute values over all attribute classes and all clusters in a network.

Alternatively, the attribute’s $Q$ underexpression can also be tested.  
Here, we want to see if the number of elements in the cluster with a specific attribute value is significantly lower than randomly selecting the elements from the total system elements.
The probability under the null hypothesis that the value of an attribute $Q$ in a cluster $C$ is smaller than the observed value in the system can be obtained from the left tail of the hypergeometric distribution, as follows:
\begin{equation}
p_{u}(N_{C,Q}) = \sum_{X=0}^{N_{C,Q}}H(X|N, N_C, N_Q)
\label{eq:hyper_under}
\end{equation}
Again, if $p_{u}(N_{C,Q})$ is smaller than a given FDR-adjusted  p-threshold, we say that the attribute $Q$ is underexpressed in cluster $C$. 
We used the same setting for the FDR correction.

\section*{Results}
\begin{figure}[h]
\centering
\captionsetup[subfigure]{justification=centering}
\begin{multicols}{4}
\begin{subfigure}{0.25\textwidth}
\centering
\includegraphics[width=1\textwidth,height=0.4\textheight,keepaspectratio]{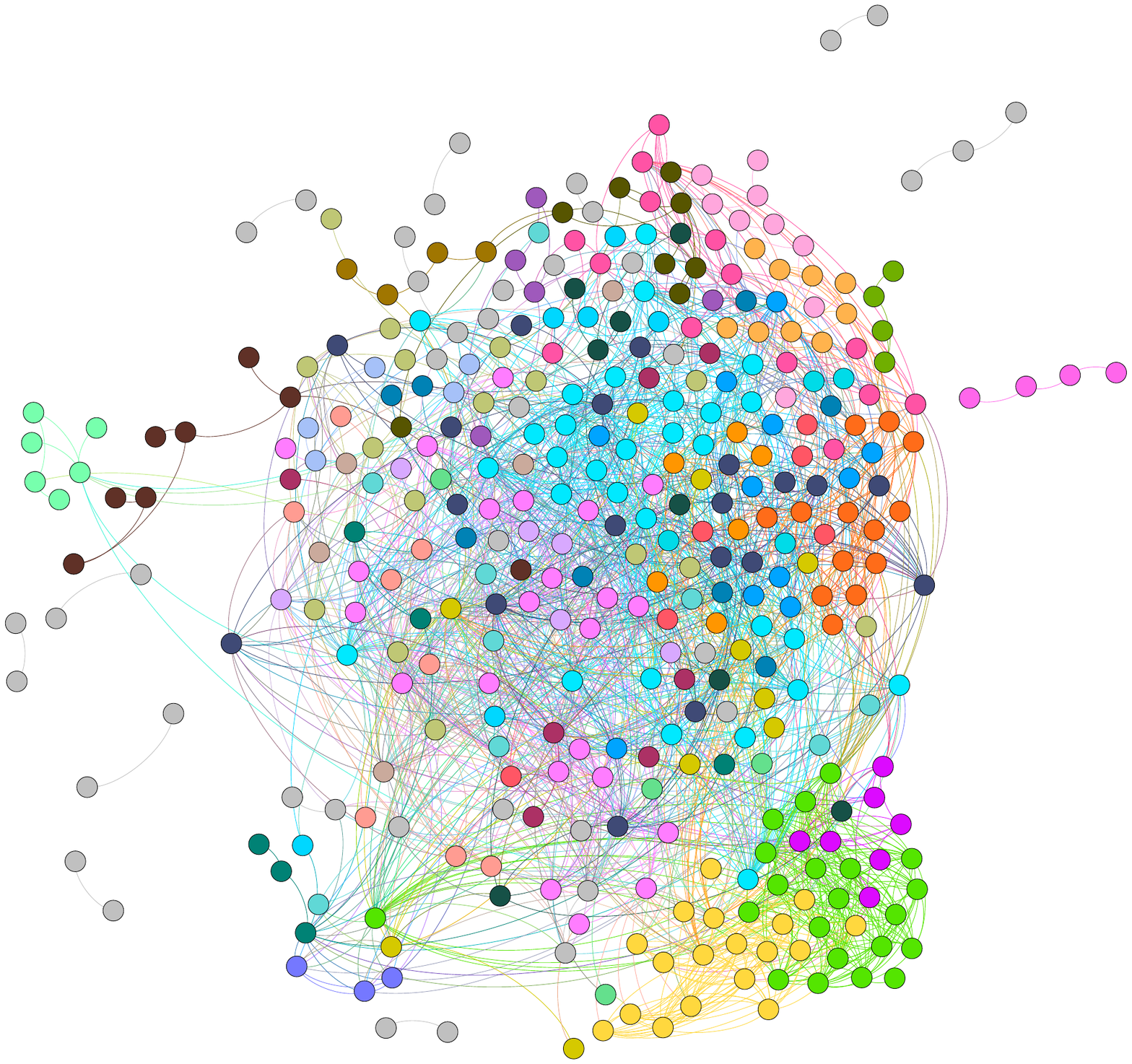}\par 
\caption{}
\end{subfigure}%
\begin{subfigure}{0.25\textwidth}
\centering
\includegraphics[width=1\textwidth,height=0.4\textheight,keepaspectratio]{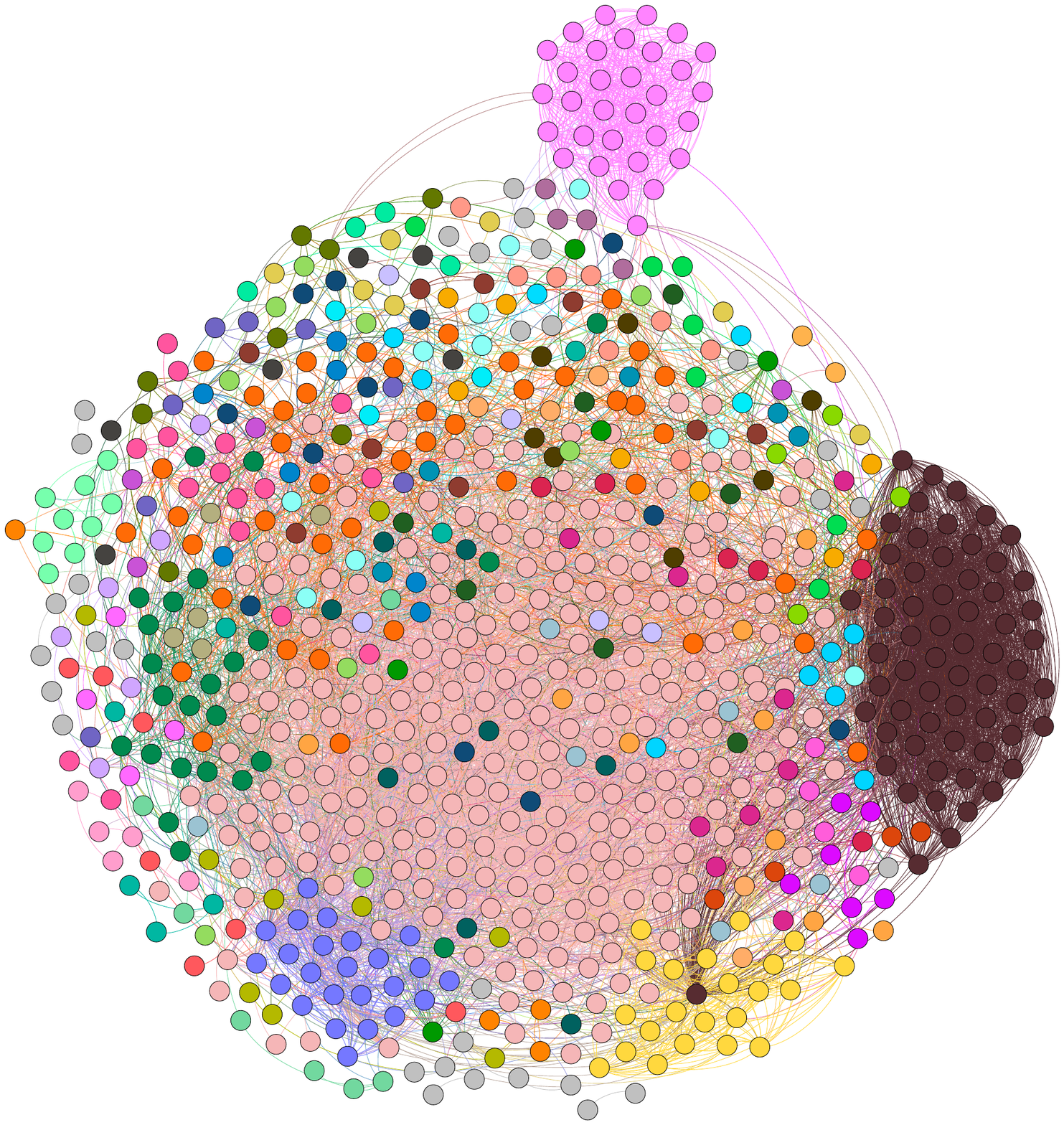}\par
\caption{}
\end{subfigure}%
\begin{subfigure}{0.25\textwidth}
\centering
\includegraphics[width=1\textwidth,height=0.4\textheight,keepaspectratio]{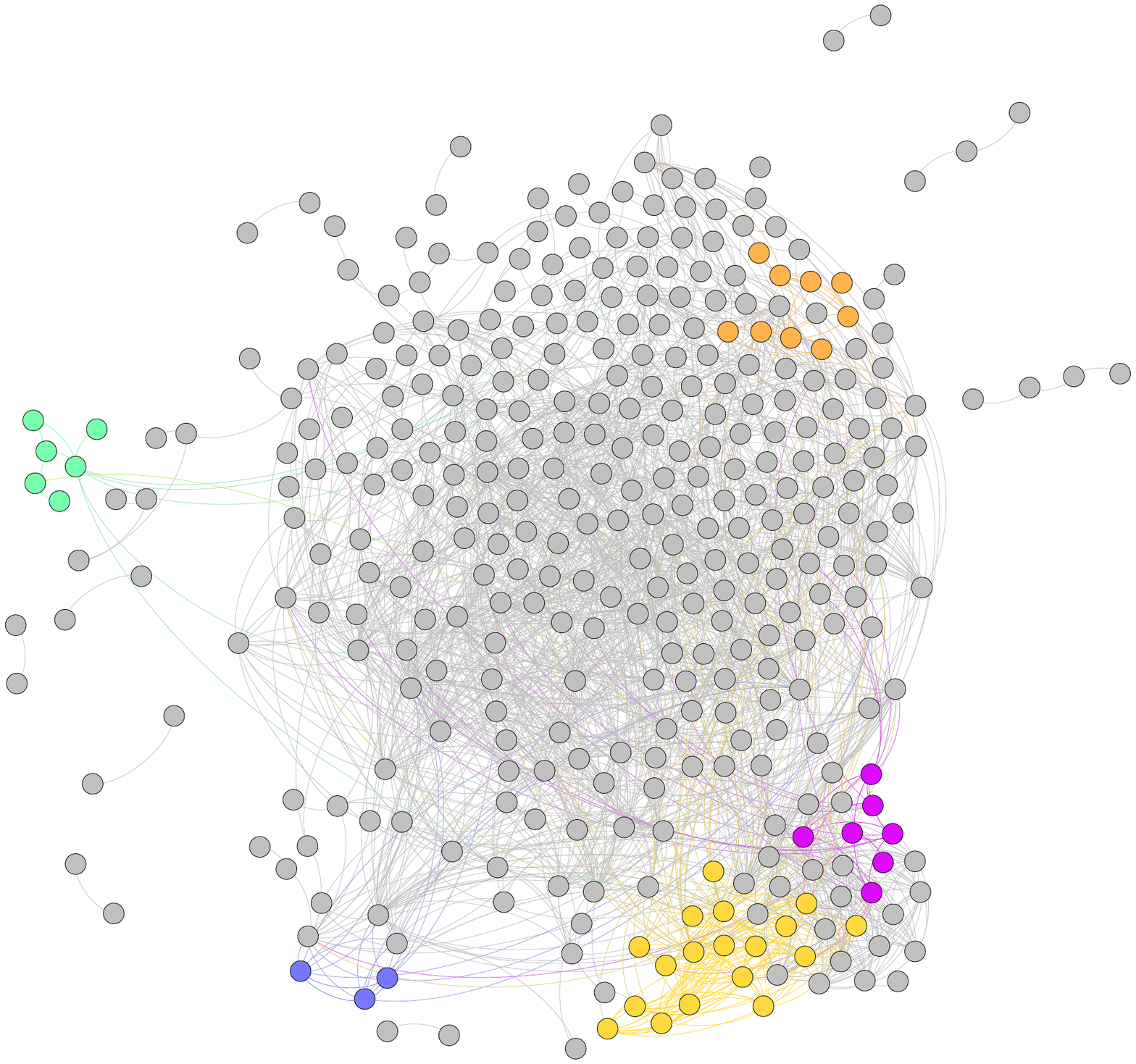}\par 
\caption{}
\end{subfigure}%
\begin{subfigure}{0.25\textwidth}
\centering
\includegraphics[width=1\textwidth,height=0.4\textheight,keepaspectratio]{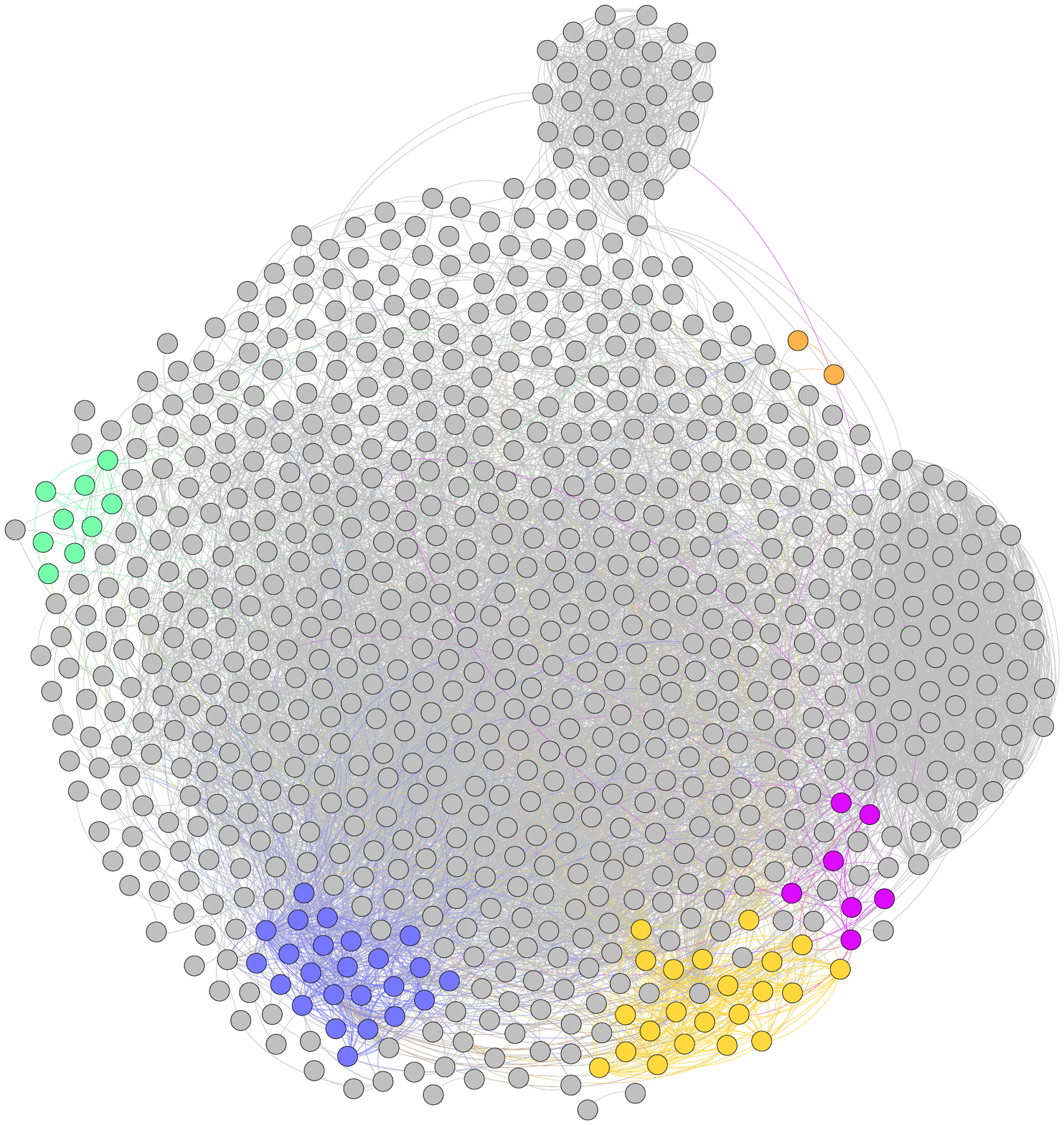}\par 
\caption{}
\end{subfigure}%
\end{multicols}
\caption {
{\bf Infomap clusters and their evolution for Kemira GrowHow (FI0009012843).}
Community detection is used with weighted links based on the total number of buy state, sell state, and day trade link types between two investors.
(a) FDR: 54 clusters, first year after IPO, 
(b) FDR: 64 clusters, second year after IPO,
(c) and (d) show five statistically significant overlapping clusters in both years.
Node position is fixed. 
The colours of reoccurring clusters in all graphs coincide. 
In (a) and (b), each cluster has a unique colour, with the exception of those with fewer than four elements, which are coloured in grey.
\label{fig:netw_infomap_com}
}
\end{figure}

Using the SVN methodology, for each of the 69 securities we infer $b$, $s$ and $bs$ trading state networks for the first and the second year after their IPO dates. 
In order to identify investor clusters  we start by aggregating the networks for all three possible joint-trading states into one weighted network.
Each link in the network is given the weight $w\in \{1,2,3\}$ depending on how many validated trading states have been observed for a given investor pair\footnote{For example, if the given investors were timing their buy transactions similarly so that they have a statistically validated link in the buying state but there were no statistical association with the sell and buy–sell states, then the weight of the link between the investors would be 1.}.
Finally, for each weighted network we identify clusters using Infomap community detection algorithm\footnote{We use \textit{igraph} implementation of the Infomap algorithm with $100$ as the parameter for the number of trials.} \citep{rosvall2008maps}.  
Identified communities are locally dense connected subgraphs in a network that play an important role in understanding a system’s topology.  
In the current paper, communities represent investor clusters that are timing their trades synchronously throughout the year. 
Table \ref{tab:number_clusters_all} summarizes the number of observed clusters during the first and the second year.
For example, during the first year, $54$ investor clusters were identified in the securities' Kemira GrowHow (FI0009012843) networks, while during the second year $64$ clusters were formed.
Fig. \ref{fig:netw_infomap_com} (a) and (b) visualise the later Infomap clusters for the first- and second-year networks.

\begin{table}[htbp]
\caption{
\textbf{Investor network clusters' statistics.}
Columns ‘Unique clusters Y1 (Y2)’ show the number of asset-specific investor clusters over all clusters observed in the first (second) year networks.
Here, asset-specific investor clusters are defined as those that were not observed in other IPO networks.
The number in the brackets () shows the ratio in percentage.
'Persisting clusters Y1 $ \rightarrow $ Y2' shows the number of clusters with statistically significant overlaps in the first and the second years. 
Note that cluster split and merge, and thus the number of persisted clusters is not necessarily the same for both years.
Columns 'Unique investors Y1 (Y2)' show the total number of investors per ISIN in a year.
Columns 'Active investors Y1 (Y2)' show the total number of investors who traded at least 5 d. per ISIN in a year.
The column 'Active investors $Y1 \cap Y2$' shows the total number of investors who traded at least 5 d. per ISIN in both first and second year after IPO.
The column 'Median cluster size' shows the number of investors in a median-sized cluster in the first (second) year network.
ISINs from the error-free set are marked in bold.
\label{tab:number_clusters_all}
}

\centering
\resizebox{!}{0.4\textheight}{
\begin{tabular}{lcccrrrrrrr}
\hline
ISIN &    \begin{tabular}{c} Unique \\
clusters \\ Y1  \end{tabular} &    \begin{tabular}{c} Unique \\
clusters \\ Y2  \end{tabular} & 
\begin{tabular}{c} Persisting \\
 clusters \\ Y1 $ \rightarrow $ Y2 \end{tabular} & \begin{tabular}{c} Unique \\
investors \\ Y1  \end{tabular} & \begin{tabular}{c} Active \\
investors \\ Y1  \end{tabular} & \begin{tabular}{c} Unique \\
investors \\ Y2  \end{tabular} & \begin{tabular}{c} Active \\
investors \\ Y2 \end{tabular} & \begin{tabular}{c} Active \\
investors \\ $Y1 \cap Y2$  \end{tabular} & \begin{tabular}{c} Median \\
 cluster \\ size\end{tabular}\\
\thickhline
\textbf{FI0009004881} & 1/14 (7\%) & 1/10 (10\%) &    &    714 &   107 &    875 &    111 &     38 &  3 (2) \\
FI0009800346 & 0/28 (0\%) & 4/40 (10\%) & $10 \rightarrow 10$ &   3747 &   251 &   4410 &    316 &    117 &  3 (4) \\
FI0009800320 & 1/8 (13\%) & 0/21 (0\%) & $3 \rightarrow 3$ &   1867 &   106 &   2365 &    146 &     48  &  4 (3) \\
FI0009900336 & 0/0 (0\%) & 0/11 (0\%) &    &    441 &    22 &   1465 &     90 &     10 &  0 (4) \\
FI0009005318 & 0/5 (0\%) & 0/13 (0\%) & $1 \rightarrow 1$ &    545 &    51 &    653 &     74 &     19 &  2 (2) \\
FI0009800643 & 0/5 (0\%) & 0/32 (0\%) & $1 \rightarrow 1$ &    536 &    50 &   2730 &    261 &     23 &  3 (3) \\
FI0009005870 & 0/6 (0\%) & 0/14 (0\%) & $1 \rightarrow 1$ &    947 &    68 &    734 &     67 &     22  &  2 (2) \\
FI0009005953 & 1/14 (7\%) & 0/10 (0\%) & $1 \rightarrow 1$ &   2108 &   131 &   2509 &    104 &     46 &  2 (2) \\
FI0009005961 & 0/30 (0\%) & 4/65 (6\%) & $11 \rightarrow 13$ &   3570 &   280 &   5555 &    501 &    159 &  3 (4) \\
FI0009005987 & 8/82 (10\%) & 11/110 (10\%) & $29 \rightarrow 32$ &  11093 &   678 &  15139 &    906 &    314 &  3 (4) \\
FI0009006381 & 0/32 (0\%) & 0/38 (0\%) & $2 \rightarrow 2$ &   5277 &   226 &   4085 &    316 &     74 &  3 (3) \\
FI0009006415 & 0/11 (0\%) & 0/5 (0\%) &   &   1258 &    85 &    601 &     35 &      9 &  3 (2) \\
FI0009005805 & 1/39 (3\%) & 1/12 (8\%) &    &   4835 &   294 &   1560 &    102 &     42 &  3 (3) \\
FI0009006589 & 0/11 (0\%) & 0/1 (0\%) &    &   1853 &    92 &    274 &     20 &      7 &  3 (2) \\
FI0009006621 & 2/66 (3\%) & 1/63 (2\%) & $6 \rightarrow 7$ &  14372 &   469 &  11033 &    565 &    155 &  4 (5) \\
FI0009006738 & 0/38 (0\%) & 3/71 (4\%) & $2 \rightarrow 2$ &   5789 &   305 &   7261 &    542 &     84 &  2 (3) \\
FI0009006696 & 0/5 (0\%) & 0/5 (0\%) & $1 \rightarrow 1$ &   1073 &    72 &    672 &     39 &     15 &  3 (3) \\
FI0009006761 & 0/8 (0\%) & 0/7 (0\%) &    &   1000 &    52 &   1252 &     55 &     17 &  4 (3) \\
FI0009007017 & 0/0 (0\%) & 1/4 (25\%) &    &    534 &    29 &    888 &     38 &      8 &  0 (2) \\
FI0009007025 & 0/9 (0\%) & 0/20 (0\%) & $1 \rightarrow 1$ &   1025 &    68 &   1951 &    125 &     29 &  2 (2) \\
FI0009007066 & 0/8 (0\%) & 0/3 (0\%) &    &   1984 &    60 &    341 &     28 &     14 &  3 (2) \\
FI0009006829 & 0/5 (0\%) & 1/15 (7\%) & $1 \rightarrow 1$ &   1902 &    56 &   3212 &    136 &     27 &  2 (3) \\
FI0009007215 & 3/10 (30\%) & 1/25 (4\%) & $1 \rightarrow 1$ &   1673 &    83 &   2952 &    156 &     21 &  5 (2) \\
FI0009007264 & 7/113 (6\%) & 52/475 (11\%) & $68 \rightarrow 98$ &   8067 &   854 &  43745 &   4288 &    482 &  3 (4) \\
FI0009007371 & 20/272 (7\%) & 136/818 (17\%) & $227 \rightarrow 389$ &  33419 &  2633 &  82702 &  10050 &   1467 &  6 (7) \\
FI0009007355 & 0/1 (0\%) & 0/0 (0\%) &    &    747 &    13 &    774 &     32 &      6 &  2 (0) \\
FI0009007132 & 8/111 (7\%) & 0/54 (0\%) & $5 \rightarrow 6$ &  22617 &   943 &  18156 &    514 &    218 &  5 (6) \\
FI0009007629 & 0/2 (0\%) & 0/12 (0\%) &    &    596 &    53 &   1426 &     91 &     14 &  2 (3) \\
FI0009801286 & 0/15 (0\%) & 0/1 (0\%) &    & 3191 &   115 &    968 &     38 &     16 &  4 (2) \\
FI0009007553 & 6/81 (7\%) & 2/99 (2\%) & $17 \rightarrow 16$ &   9492 &   657 &   9449 &    997 &    182 &  4 (3) \\
FI0009007728 & 1/44 (2\%) & 1/31 (3\%) & $1 \rightarrow 1$ &   7219 &   303 &   3355 &    322 &     76 &  4 (2) \\
FI0009007546 & 0/1 (0\%) & 0/0 (0\%) &    &    232 &    26 &     97 &      7 &      5 &  2 (0) \\
FI0009007686 & 2/5 (40\%) & 0/1 (0\%) &    &    753 &    45 &    417 &     22 &      3 &  2 (2) \\
FI0009007694 & 4/27 (15\%) & 0/2 (0\%) &    &   2774 &   176 &   1909 &     88 &     28 &  3 (2) \\
FI0009006886 & 2/11 (18\%) & 0/2 (0\%) &    &   1849 &   103 &    819 &     39 &     11 &  2 (2) \\
FI0009007819 & 4/135 (3\%) & 2/123 (2\%) & $13 \rightarrow 14$ &  16608 &  1223 &   8287 &   1117 &    329 &  6 (5) \\
FI0009007835 & 2/41 (5\%) & 0/34 (0\%) & $4 \rightarrow 4$ &   6320 &   283 &   3910 &    235 &     85  &  3 (3) \\
FI0009007884 & 11/136 (8\%) & 2/100 (2\%) & $4 \rightarrow 3$ &  58326 &  1049 &  20940 &    934 &    277 &  3 (5) \\
FI0009008080 & 1/11 (9\%) & 0/9 (0\%) & $1 \rightarrow 1$ &   1296 &    91 &   1094 &    102 &     27 &  2 (3) \\
FI0009007918 & 1/87 (1\%) & 1/79 (1\%) & $9 \rightarrow 10$ &   7136 &   802 &   7199 &   1051 &    256 &  6 (6) \\
FI0009801310 & 15/169 (9\%) & 4/218 (2\%) & $79 \rightarrow 84$ & 30706 &  2328 &  20898 &   2497 &    672 &  2 (7) \\
FI0009008221 & 38/337 (11\%) & 8/252 (3\%) & $198 \rightarrow 172$ &  35617 &  3454 &  17235 &   2541 &    985 &  7 (6) \\
FI0009902530 & 6/62 (10\%) & 2/65 (3\%) & $8 \rightarrow 8$ &  25808 &   572 &  12223 &    614 &    200 &  6 (5) \\
FI0009008924 & 0/12 (0\%) & 1/2 (50\%) &    &   2644 &   151 &   1070 &     74 &     27 &  4 (2) \\
FI0009008833 & 0/2 (0\%) & 0/3 (0\%) &    &    674 &    35 &    615 &     45 &     14  &  2 (2) \\
FI0009009146 & 2/28 (7\%) & 0/1 (0\%) & $1 \rightarrow 1$ &   3444 &   285 &   1188 &     70 &     34 &  2 (2) \\
FI0009009054 & 1/8 (13\%) & 1/8 (13\%) &    &   1832 &   135 &   1120 &    112 &     34  &  2 (2) \\
FI0009009633 & 4/19 (21\%) & 1/14 (7\%) &     &   2847 &   224 &   1771 &    107 &     33 &  2 (3) \\
FI0009009567 & 4/12 (33\%) & 0/8 (0\%) &    &   1614 &   112 &   1380 &     96 &     28 &  2 (2) \\
FI0009008270 & 6/53 (11\%) & 0/12 (0\%) &    &   5743 &   442 &   2558 &    137 &     64 &  3 (2) \\
FI0009009674 & 1/24 (4\%) & 1/13 (8\%) & $4 \rightarrow 4$ &   2033 &   226 &   3746 &    161 &     88 &  4 (6) \\
FI0009009377 & 1/5 (20\%) & 1/6 (17\%) &    &   2779 &   151 &   2329 &    133 &     26 &  2 (2) \\
FI0009010219 & 1/8 (13\%) & 1/5 (20\%) &    &   1438 &    84 &   1078 &     66 &     19 &  4 (2) \\
FI0009010854 & 0/2 (0\%) & 0/14 (0\%) &    &    573 &    41 &   1164 &    114 &     19 &  4 (4) \\
FI0009010862 & 0/5 (0\%) & 1/10 (10\%) &    &    879 &    66 &   1604 &     99 &     16 &  2 (4) \\
SE0000667925 & 22/120 (18\%) & 7/129 (5\%) & $8 \rightarrow 9$ &  17759 &  1186 &  21725 &   1580 &    476 &  4 (7) \\
\textbf{SE0000110165} & 1/4 (25\%) & 0/0 (0\%) &    &    576 &    43 &    176 &      9 &      3 &  2 (0) \\
\textbf{FI0009012843} & 5/54 (9\%) & 2/64 (3\%) & $5 \rightarrow 5$ &   8047 &   464 &   9609 &    818 & 183 & 5 (6) \\
\textbf{FI0009013296} & 33/262 (13\%) & 42/336 (13\%) & $180 \rightarrow 221$ &  24350 &  3518 &  22421 &   3603 &   1555 &  7 (7) \\
\textbf{FI0009013429} & 12/133 (9\%) & 5/89 (6\%) & $26 \rightarrow 24$ &   9945 &  1016 &   6012 &    691 &    326 &  3 (3) \\
\textbf{FI0009013312} & 3/26 (12\%) & 0/13 (0\%) & $2 \rightarrow 1$ &   2667 &   224 &   1204 &    125 &     52 &  4 (4) \\
\textbf{FI0009013403} & 11/112 (10\%) & 4/92 (4\%) & $45 \rightarrow 44$ &  10234 &  1084 &   7952 &    769 &    409 &  2 (3) \\
\textbf{FI0009013924} & 3/29 (10\%) & 1/11 (9\%) &   &   1804 &   192 &   2104 &    235 &     43 &  2 (2) \\
\textbf{FI0009010391} & 3/37 (8\%) & 6/50 (12\%) & $3 \rightarrow 3$ &   8822 &   306 &   5915 &    434 &    114 &  3 (5) \\
\textbf{FI0009013593} & 3/17 (18\%) & 0/0 (0\%) &   & 2345 &   162 &    870 &      9 &      4 &  3 (0) \\
\textbf{FI0009014344} & 2/3 (67\%) & 0/0 (0\%) &    &   2815 &    83 &   1128 &     34 &     15  &  2 (0) \\
\textbf{FI0009014351} & 11/56 (20\%) & 0/6 (0\%) &    &  10338 &   399 &   3267 &    135 &     69 &  2 (2) \\
\textbf{FI0009012413} & 11/22 (50\%) & 5/22 (23\%) & $1 \rightarrow 1$ &   4788 &   243 &   6627 &    237 &     62 &  2 (5) \\
\textbf{FI0009015309} & 9/64 (14\%) & 3/29 (10\%) & $1 \rightarrow 1$ &   6748 &   521 &   2208 &    187 &     95 &  3 (4) \\
\hline
\end{tabular}}
\end{table}

Next, for each security, we detect clusters with a statistically significant investor overlap between the first and second year.
The summary of statistically validated cluster time persistence for all 69 securities is presented in the fourth column of Table \ref{tab:number_clusters_all}.
For example, in the Kemira GrowHow networks, only $5$ of the $54$, i.e. $9\%$ of clusters identified in the first year were observed in the second year.
Fig. \ref{fig:netw_infomap_com} (c) and (d) display those five clusters that persisted over the first two years after the IPO.
The observation in the example that only a small number of clusters persist into the second year is consistent for the majority of the analysed IPO securities.
However, there are several securities for which more than a half of the first year clusters persist into the following year.
A sample of time persistent clusters and their composition in terms of investor attributes are visualised in the Appendix Fig. A.1 and A.2.

By calculating the fraction of clusters that do not persist into the second year, we observe that over all 69 securities on average $88\%$ of the first-year clusters are not observed in the following year while the same number falls to $78\%$ for mature company networks inferred during the same periods (more details about the comparison to mature companies are provided in the following section). This observation can suggest the existence of IPO trading strategy related clusters that form exclusively during the first year after the IPO date and break up in the following year.

\begin{figure}[!h]
    \centering
      \includegraphics[width=0.5\textwidth,keepaspectratio]{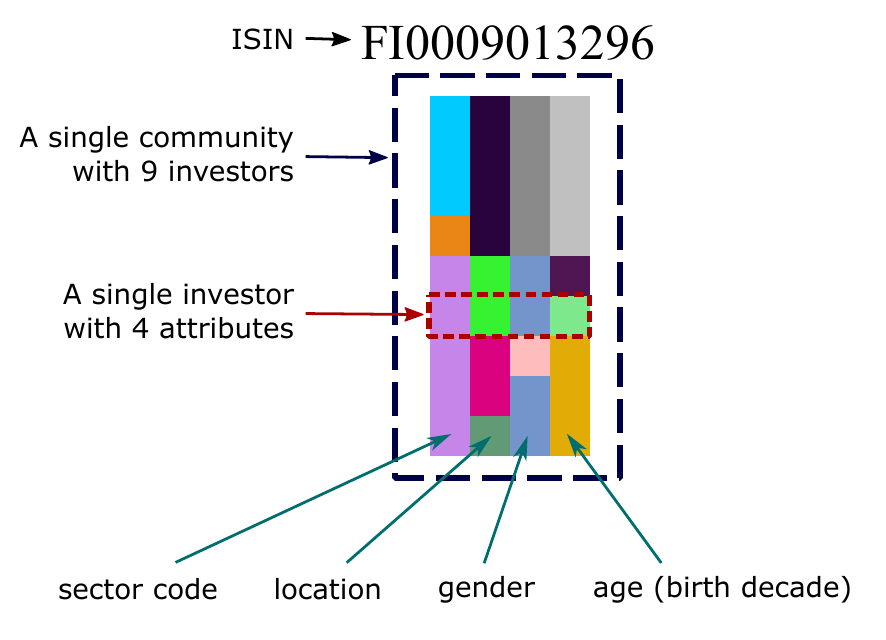}
    
  \caption {
  {\bf Graphical representation of the clusters.}
A single cluster is visualised as a rectangle block, where a row represents one investor with four attributes: sector code, location, gender and birth year decade.
  \label{fig:com_expl}  
  \\
\textbf{Sector code:}
\protect\tikz{\protect\draw [black, thin, fill=cc686e9] (4pt,4pt) rectangle (0.4,0.4) ;} - Households, 
\protect\tikz{\protect\draw [black, thin, fill=c5fc613] (4pt,4pt) rectangle (0.4,0.4);} - Non-financial, 
\protect\tikz{\protect\draw [black, thin, fill=c00caff] (4pt,4pt) rectangle (0.4,0.4);} - Financial-Insurance, 
\protect\tikz{\protect\draw [black, thin, fill=cea8615] (4pt,4pt) rectangle (0.4,0.4);} - General-Government, 
\protect\tikz{\protect\draw [black, thin, fill=c2e9072] (4pt,4pt) rectangle (0.4,0.4);} - Non-Profit, 
\protect\tikz{\protect\draw [black, thin, fill=cff5c81] (4pt,4pt) rectangle (0.4,0.4);} - Rest-World. \\
\textbf{Geographic location:}
\protect\tikz{\protect\draw [black, thin, fill=cda027f] (4pt,4pt) rectangle (0.4,0.4) ;} - Helsinki, 
\protect\tikz{\protect\draw [black, thin, fill=c29033c] (4pt,4pt) rectangle (0.4,0.4);} - South-West, 
\protect\tikz{\protect\draw [black, thin, fill=c639a76] (4pt,4pt) rectangle (0.4,0.4);} - Western-Tavastia, 
\protect\tikz{\protect\draw [black, thin, fill=c36f230] (4pt,4pt) rectangle (0.4,0.4);} - Central-Finland, 
\protect\tikz{\protect\draw [black, thin, fill=cf9f99d] (4pt,4pt) rectangle (0.4,0.4);} - Northern-Finland, 
\protect\tikz{\protect\draw [black, thin, fill=cbae396] (4pt,4pt) rectangle (0.4,0.4);} - Ostrobothnia,
\protect\tikz{\protect\draw [black, thin, fill=c9ee0e0] (4pt,4pt) rectangle (0.4,0.4) ;} - Rest-Uusimaa, 
\protect\tikz{\protect\draw [black, thin, fill=ca5bdcc] (4pt,4pt) rectangle (0.4,0.4);} - Eastern-Tavastia, 
\protect\tikz{\protect\draw [black, thin, fill=c4cdc38] (4pt,4pt) rectangle (0.4,0.4) ;} - Eastern-Finland, 
\protect\tikz{\protect\draw [black, thin, fill=cd340bd] (4pt,4pt) rectangle (0.4,0.4);} - South-East, 
\protect\tikz{\protect\draw [black, thin, fill=c874010] (4pt,4pt) rectangle (0.4,0.4);} - Northern-Savonia. \\
\textbf{Gender:}
\protect\tikz{\protect\draw [black, thin, fill=c7495cb] (4pt,4pt) rectangle (0.4,0.4) ;} - Male, 
\protect\tikz{\protect\draw [black, thin, fill=cffbcbc] (4pt,4pt) rectangle (0.4,0.4);} - Female, 
\protect\tikz{\protect\draw [black, thin, fill=c8a8a8a] (4pt,4pt) rectangle (0.4,0.4);} - No-Gender. \\
\textbf{Decade:}
\protect\tikz{\protect\draw [black, thin, fill=cc0c0c0] (4pt,4pt) rectangle (0.4,0.4) ;} - No-Age, 
\protect\tikz{\protect\draw [black, thin, fill=c4b771a] (4pt,4pt) rectangle (0.4,0.4);} - 1910, 
\protect\tikz{\protect\draw [black, thin, fill=cda0c5a] (4pt,4pt) rectangle (0.4,0.4);} - 1920,
\protect\tikz{\protect\draw [black, thin, fill=c7e3f9c] (4pt,4pt) rectangle (0.4,0.4) ;} - 1930, 
\protect\tikz{\protect\draw [black, thin, fill=c501653] (4pt,4pt) rectangle (0.4,0.4);} - 1940, 
\protect\tikz{\protect\draw [black, thin, fill=c7de88c] (4pt,4pt) rectangle (0.4,0.4);} - 1950,
\protect\tikz{\protect\draw [black, thin, fill=ce1ac06] (4pt,4pt) rectangle (0.4,0.4) ;} - 1960, 
\protect\tikz{\protect\draw [black, thin, fill=c430337] (4pt,4pt) rectangle (0.4,0.4);} - 1970, 
\protect\tikz{\protect\draw [black, thin, fill=c6e4b83] (4pt,4pt) rectangle (0.4,0.4);} - 1980,
\protect\tikz{\protect\draw [black, thin, fill=ce40030] (4pt,4pt) rectangle (0.4,0.4);} - 1990, 
\protect\tikz{\protect\draw [black, thin, fill=c102ee2] (4pt,4pt) rectangle (0.4,0.4);} - 2000.}
\end{figure}

Additionally, we analyse cluster overlap across multiple securities, separately for the first- and second-year networks.
The second and third columns in Table \ref{tab:number_clusters_all} show the number of asset-specific clusters over the total number of communities in the first and second year.
Here, by asset-specific clusters, we refer to the clusters that are not observable within investor networks of the same year for other IPO securities in our investigated 69 security universe.
The number of observed asset-specific clusters is rather small and is around $15\%$ ($9\%$) during the first (second) year averaged over all 69 securities. This means that the majority of investor clusters are found to be present in multiple securities, i.e. they execute synchronized trading strategies over multiple IPOs. Note that this cluster synchronization is observed even though the network inference periods are not aligned in time. The observed decrease in the overall percentage of asset-specific clusters hints that during the second year after IPO more clusters use non-IPO related trading strategies. This is later supported by the mature security analysis (see the next section and Table \ref{tab:overlaps_with_mature}).
Fig. A.3 in Appendix A shows a sample of clusters with statistically significant investor overlap across multiple securities.

\begin{figure}[!ht]
    \centering
      \includegraphics[width=\textwidth,keepaspectratio]{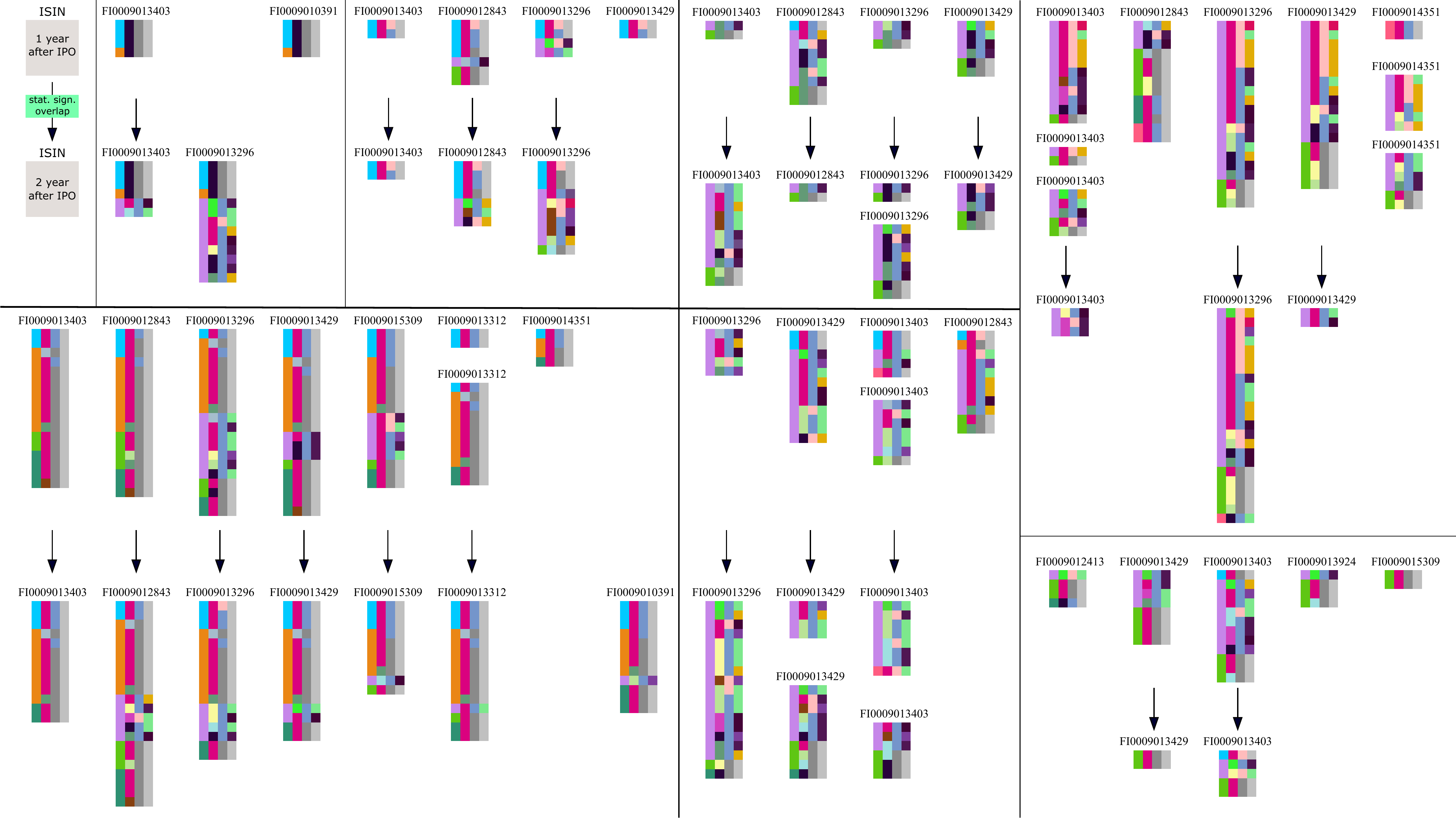}
    
  \caption {
  {\bf Statistically significant cluster overlaps across multiple securities and over time.}
  \label{fig:comb_evol1}  
  The figure contains many subfigures separated by borders. Each subfigure presents a cluster of investors that spans over multiple securities and persists in time.
  The row alignment shows statistically similar clusters in the same year: the top row is the first after the IPO, and the bottom row is the second year after the IPO. 
  The downward arrows show the cluster timewise evolution from the first to the second year for the same security. 
  A cluster is represented by the rectangle. 
  Each cluster is composed of investors with four attributes: sector code, geographic location, gender and decade.
  See the attribute colour mapping in Fig. \ref{fig:com_expl}.
}
\end{figure}

Combining the previous results together, we observe persistent clusters that emerge in investor networks over multiple securities. 
Fig. \ref{fig:com_expl} explains the visualisation of a cluster in this study and Fig. \ref{fig:comb_evol1} shows a sample of clusters that both, overlap over time and over multiple securities. 
In the figure, the top (bottom) row of the group refers to the first- (second-) year clusters. 
Moreover, the downward arrows associate statistically similar clusters in the first- and second-year networks. 
The arrows between the clusters in the same year after IPO are omitted for the simplification of the visualisation.
Notably, even if some of the clusters are not persistent over time, quite often they appear over different securities.

Next, we  analyse the over- and underexpression of the investor attributes in the identified investor clusters. We say that a cluster is overexpressing (underexpressing) an attribute if the number of investors in the cluster with that particular attribute is significantly higher (lower) than could be expected under the null model defined in the Methods section. We are primarily interested in the sector code attribute analysis, where investors can be assigned households, nonfinancial corporations, financial and insurance corporations, government, nonprofit institutions, and the rest of the world attribute. Additionally, we test whether or not attributes related to gender, age or geographical location are over- or underexpressed\footnote{No-Gender and No-Age attributes refer to the institutional investors, but also to the individual investors who had no gender and/or birth year indicated in the data.}.

\begin{figure}
    \centering
    \includegraphics[width=\textwidth]{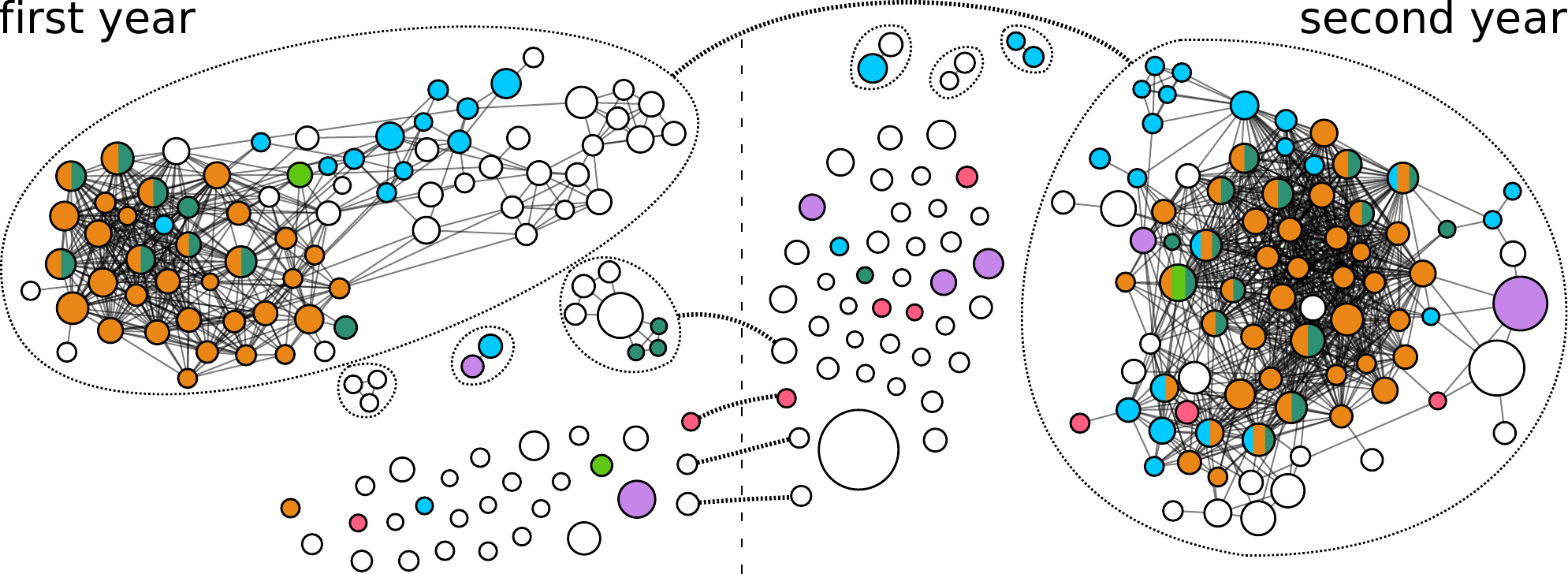}
    \caption{\textbf{Network of investor clusters with overexpressed attributes.} On the left-hand-side are the clusters observed in the first year after respective IPOs and right-hand-side, in the second year. Investor cluster nodes are connected with continuous links if they share statistically significant number of individual investors. Dashed links represent statistical similarity between some of the connected cluster components in the first and the second year after the IPOs. Node colours identify overexpressed sector codes within clusters. For overexpressed geographical location see Appendix Fig. B.1, for underexpressed attributes see Fig. \ref{fig:under_expressing_clusters} and for all overexpressed and underexpressed attributes see Appendix Tables B.1 and B.2.
    \textbf{Sector code:}
    \protect\tikz{\protect\draw [black, thin, fill=cc686e9] (4pt,4pt) circle (0.13);} - Households, 
    \protect\tikz{\protect\draw [black, thin, fill=c5fc613] (4pt,4pt) circle (0.13);} - Non-financial, 
    \protect\tikz{\protect\draw [black, thin, fill=c00caff] (4pt,4pt) circle (0.13);} - Financial-Insurance, 
    \protect\tikz{\protect\draw [black, thin, fill=cea8615] (4pt,4pt) circle (0.13);} - General-Government, 
    \protect\tikz{\protect\draw [black, thin, fill=c2e9072] (4pt,4pt) circle (0.13);} - Non-Profit, 
    \protect\tikz{\protect\draw [black, thin, fill=cff5c81] (4pt,4pt) circle (0.13);} - Rest-World. \\}
    \label{fig:sector_code_over_expressing_clusters}
\end{figure}

Over all 69 securities, we identify $115$ ($28$) investor clusters with $182$ ($40$) overexpressed (underexpressed) attributes during the first year after the IPO, and $130$ ($44$) investor clusters with $236$ ($70$) overexpressed (underexpressed) attributes during the second year. The number of overexpressed (underexpressed) attributes is larger than the number of investor clusters, because each cluster can overexpress (underexpress) more than one attribute. The overexpressed clusters are observed over $28$ different securities during the first year after IPO and for $27$ different securities during the second year after IPO. As for the underexpressed clusters, they are observed over $16$ securities during the first year and $20$ securities during the second year after IPO.\\
\\
In order to present the attribute analysis in a concise way, we use the fact that the same clusters appear over multiple securities and assign overexpressed (underexpressed) investor clusters into groups if they are statistically similar. Fig. \ref{fig:sector_code_over_expressing_clusters} presents the resulting sector code attribute overexpressing investor cluster networks for the first and second years after respective IPOs. In the figure, nodes on the left (right) hand side of the vertical dashed line represent investor clusters observed in the first (second) year after IPO. Statistically similar cluster nodes are connected with links and dotted lines circle network components. Each connected component in the network relates to a group of clusters with a statistically similar investor composition. The dashed lines crossing from the left to the right-hand-side indicate that there is a statistical similarity for some of the clusters in the components between the first and the second year.

\begin{figure}[H]
    \centering
    \includegraphics[width=\textwidth]{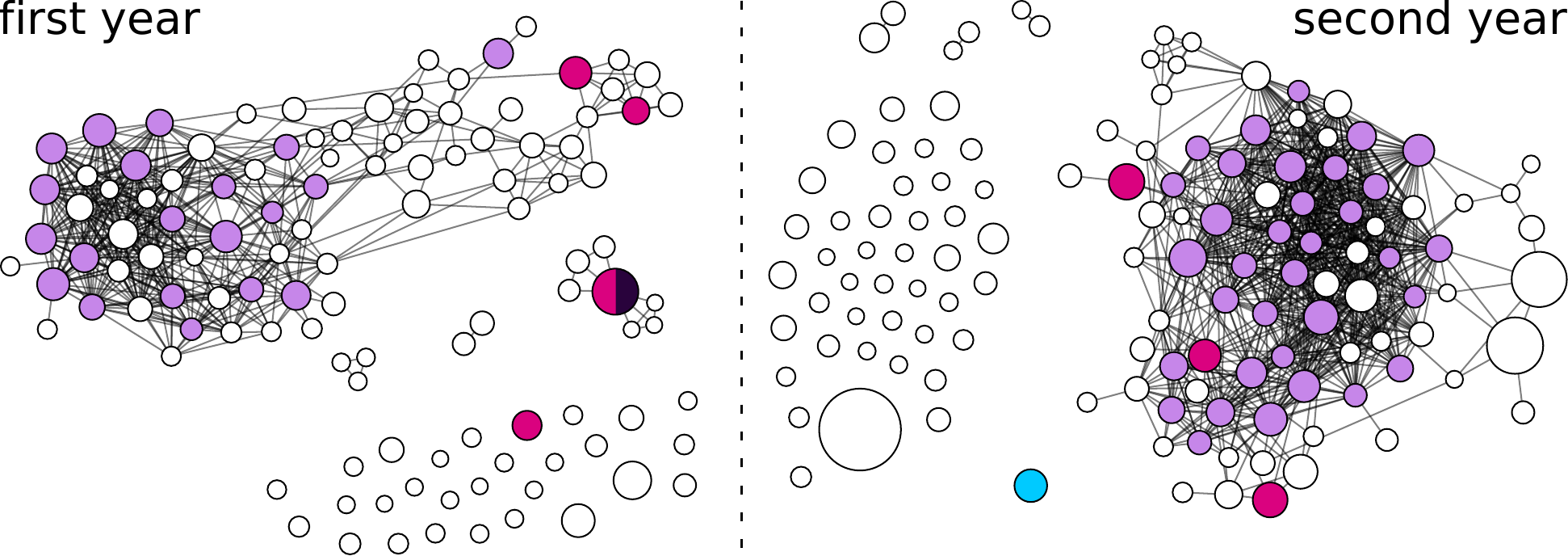}
    \caption{\textbf{Network of investor clusters with underexpressed attributes.} On the left-hand-side are the clusters observed in the first year after respective IPOs and right-hand-side, in the second year. Investor cluster nodes are connected with continuous links if they share statistically significant number of individual investors. 
    Node colours identify underexpressed sector code and geographical location attributes within clusters.
\textbf{Sector code:}
\protect\tikz{\protect\draw [black, thin, fill=cc686e9] (4pt,4pt) circle (0.13) ;} - Households, 
\protect\tikz{\protect\draw [black, thin, fill=c00caff] (4pt,4pt)  circle (0.13);} - Financial-Insurance.
\textbf{Geographic location:}
\protect\tikz{\protect\draw [black, thin, fill=cda027f] (4pt,4pt) circle (0.13);}  - Helsinki, 
\protect\tikz{\protect\draw [black, thin, fill=c29033c] (4pt,4pt) circle (0.13);} - South-West.
}
    \label{fig:under_expressing_clusters}
\end{figure}
Tables B.1 and B.2 in the Appendix summarise the over- and underexpressed cluster attributes for each investor cluster component in Fig. \ref{fig:sector_code_over_expressing_clusters} and \ref{fig:under_expressing_clusters}. 
The largest first and second year components in Fig. \ref{fig:sector_code_over_expressing_clusters} are over-represented by finance-insurance and general government institutions, as well as nonprofit organisations.
Moreover, the same components underexpress Household sector (see Fig. \ref{fig:under_expressing_clusters}), further supporting their institutional profile.
In addition, the same components overexpress location attributes, in particular Helsinki and South-West regions (see Fig. B.1 in the Appendix).  
Investor clusters with an overexpression of a geographical attribute could be observed because of some locally present investment strategy, for example an investor club, or some other means of local information transfer.
Overall, the results show that the largest cluster components mainly contain institutions that are timing their trades similarly in a year.
Compared with household investors, institutional traders form robust clusters, that execute similar trade-timing strategies over multiple IPOs, both during the first and the second year after the IPO date.
Our findings thus support the studies that provide evidence of institutional herding \citep{nofsinger1999herding, sias2004institutional}.
Some of the financial institutions, such as pension insurance companies, are driven by the same legislation and portfolio restrictions, which can lead to the same trading strategies. 
Alternatively, traders working for financial institutions have mutual and/or joint private information channels, leading to similar trade timing. 
The third explanation is that they react to public news in similar ways.

\subsection*{Do clusters of IPO investors exist with mature companies?}
To verify if our identified clusters are just IPO-related or if they exist with mature companies\footnote{Recently, the long-term evolution of the clusters of the most capitalised stock in the HSE – Nokia – has been analysed in \citet{musciotto2018long}.} as well, we compare the clusters of the new-to-the-market stocks with five mature companies (see Table \ref{tab:five_mature}).
For each mature security, just like previously for IPOs, we construct statistically validated networks and identify investor clusters with Infomap algorithm. 
When constructing the first- and second-year networks, the periods are aligned with respective IPO dates. 
This way we construct $345$ ($69 \times 5$) networks for each year.
Next, we analyse the overlaps between mature security investor clusters and the investor clusters inferred with the data from IPOs, to answer the question if the investor clusters identified with IPO securities exists with a mature company.
When statistically validating overlaps between mature and IPO security investor network clusters, we use the total number of cluster pairs with at least one investor in common between an IPO and all five mature securities as the number of tests for the FDR correction.
Table \ref{tab:overlaps_with_mature} shows the number of statistically similar clusters between the IPO and mature securities, as well as the total number of clusters observed in the IPO and the mature security during the exactly same period. Here we observe that on average over all investigated IPO securities only $16\%$ of IPO clusters are not observed in one of the five investigated mature securities during the first year after IPO, and $13\%$ during the second year. By looking at same table, we can see that only a fraction of total clusters observed in mature securities are also observed in IPO security networks. 
It can be because not all investors who trade mature securities trade recently issued securities, and if they do, not all of them might apply the same trading strategies and, therefore, not form similar synchronized clusters as in mature securities.

\begin{table}[!ht]

\centering
\caption{
{\bf Five mature companies with the highest number of transactions in HSE.}}
\begin{tabular}{clc}
\hline
ISIN  & company name & IPO date \\
\thickhline
FI0009000681 & Nokia & 1981-04-01 \\
FI0009000277 & Tieto & 1984-06-01 \\
FI0009000665 & Metsä Board B & 1987-01-02 \\
FI0009002943 & Raisio V & 1989-04-25 \\
FI0009003727 & Wärtsilä & 1991-01-17 \\
\hline
\end{tabular}
\label{tab:five_mature}
\end{table}

{
\tiny
\LTcapwidth=\textwidth
\begin{longtabu} to \textwidth {Hlllrrrrrr}
\caption{\textbf{IPO and mature companies investor clusters overlap.}
The overlap is given as A (B) \{C\}, where A is the number of the overlapping clusters in IPO stock, (B) is the number of overlapping clusters in the mature stock, and \{C\} is the total number of clusters of a mature stock.
A does not necessarily equal B because any cluster can be statistically similar to more than one cluster in another security.
'\# unique cl.' is the total number of unique IPO investor clusters over the total number of clusters, where the percentage ratio is given in brackets ().
Median and average percentage of the unique clusters over all IPO stocks are given in the bottom of the table.
}\label{tab:overlaps_with_mature}\\
\toprule
{} &      ISIN &   IPO date & year &  FI0009000681 &  FI0009003727 &  FI0009000665 &   FI0009000277 &   FI0009002943 &  \# unique cl. \\
\midrule
0   &   \multirow{2}{*}{FI0009004881} & \multirow{2}{*}{1995-01-12} &   Y1 &      2 (2) \{118\} &       1 (1) \{17\} &      2 (2) \{11\} &          \{8\} &       3 (2) \{11\} &    9/14 (64\%) \\
1   &   &	&   Y2 &     9 (11) \{142\} &       6 (8) \{33\} &      7 (9) \{34\} &        1 (1) \{5\} &       6 (6) \{57\} &    1/10 (10\%) \\
\hline
122 &   \multirow{2}{*}{FI0009800346} & \multirow{2}{*}{1995-05-11} &   Y1 &    21 (25) \{146\} &     13 (12) \{19\} &    12 (12) \{20\} &       7 (7) \{15\} &     16 (14) \{29\} &    3/28 (11\%) \\
123 &   &	&   Y2 &    26 (32) \{133\} &     18 (16) \{30\} &    23 (23) \{46\} &        1 (1) \{2\} &     15 (19) \{67\} &    6/40 (15\%) \\
\hline
120 &   \multirow{2}{*}{FI0009800320} & \multirow{2}{*}{1995-05-11} &   Y1 &      5 (7) \{146\} &       3 (3) \{19\} &      2 (3) \{20\} &       1 (2) \{15\} &       4 (5) \{29\} &     3/8 (38\%) \\
121 &   &	&   Y2 &    10 (14) \{133\} &     12 (14) \{30\} &      7 (8) \{46\} &        1 (1) \{2\} &     11 (10) \{67\} &    5/21 (24\%) \\
\hline
2   &   \multirow{2}{*}{FI0009005318} & \multirow{2}{*}{1995-06-01} &   Y1 &      3 (3) \{152\} &         \{21\} &      2 (3) \{21\} &       2 (3) \{15\} &       1 (1) \{29\} &     2/5 (40\%) \\
3   &   &	&   Y2 &     9 (10) \{123\} &       7 (8) \{34\} &      5 (6) \{38\} &        1 (1) \{3\} &       8 (9) \{62\} &     1/13 (8\%) \\
\hline
130 &   \multirow{2}{*}{FI0009900336} & \multirow{2}{*}{1995-06-01} &   Y1 &        \{152\} &         \{21\} &        \{21\} &         \{15\} &         \{29\} &      0  \\
131 &   &	&   Y2 &      6 (8) \{123\} &       6 (7) \{34\} &      5 (8) \{38\} &          \{3\} &       6 (6) \{62\} &     1/11 (9\%) \\
\hline
124 &   \multirow{2}{*}{FI0009800643} & \multirow{2}{*}{1995-09-04} &   Y1 &      5 (8) \{176\} &       1 (2) \{21\} &      2 (3) \{27\} &         \{10\} &       3 (3) \{45\} &      0/5 (0\%) \\
125 &   &	&   Y2 &    19 (31) \{133\} &     17 (19) \{38\} &    21 (22) \{38\} &        1 (1) \{2\} &     14 (15) \{67\} &    6/32 (19\%) \\
\hline
6   &   \multirow{2}{*}{FI0009005870} & \multirow{2}{*}{1996-03-27} &   Y1 &      4 (4) \{136\} &       2 (3) \{32\} &      2 (2) \{41\} &        1 (1) \{2\} &       2 (3) \{65\} &     1/6 (17\%) \\
7   &   &	&   Y2 &    13 (15) \{190\} &       6 (7) \{26\} &     10 (9) \{30\} &        1 (1) \{4\} &      9 (10) \{56\} &     0/14 (0\%) \\
\hline
12  &   \multirow{2}{*}{FI0009005987} & \multirow{2}{*}{1996-05-02} &   Y1 &    56 (54) \{128\} &     23 (24) \{30\} &    26 (24) \{43\} &          \{2\} &     33 (28) \{72\} &   18/82 (22\%) \\
13  &   &	&   Y2 &    82 (91) \{225\} &     27 (22) \{34\} &    40 (30) \{33\} &        6 (4) \{9\} &     43 (37) \{58\} &  20/110 (18\%) \\
\hline
10  &   \multirow{2}{*}{FI0009005961} & \multirow{2}{*}{1996-05-02} &   Y1 &    21 (29) \{128\} &     13 (18) \{30\} &    19 (19) \{43\} &        1 (1) \{2\} &     19 (18) \{72\} &     2/30 (7\%) \\
11  &   &	&   Y2 &    51 (73) \{225\} &     29 (26) \{34\} &    36 (26) \{33\} &        3 (3) \{9\} &     31 (33) \{58\} &    7/65 (11\%) \\
\hline
8   &   \multirow{2}{*}{FI0009005953} & \multirow{2}{*}{1996-05-02} &   Y1 &    11 (12) \{128\} &       7 (8) \{30\} &     9 (10) \{43\} &          \{2\} &       8 (9) \{72\} &    2/14 (14\%) \\
9   &   &	&   Y2 &     9 (10) \{225\} &       4 (6) \{34\} &      8 (7) \{33\} &        1 (1) \{9\} &       6 (8) \{58\} &     0/10 (0\%) \\
\hline
14  &   \multirow{2}{*}{FI0009006381} & \multirow{2}{*}{1997-04-03} &   Y1 &    26 (34) \{179\} &       5 (5) \{27\} &    14 (13) \{33\} &        2 (2) \{4\} &     19 (18) \{54\} &    4/32 (12\%) \\
15  &   &	&   Y2 &    31 (40) \{312\} &     13 (11) \{25\} &    13 (15) \{61\} &     19 (19) \{80\} &    20 (28) \{118\} &     2/38 (5\%) \\
\hline
16  &   \multirow{2}{*}{FI0009006415} & \multirow{2}{*}{1997-04-24} &   Y1 &      6 (8) \{210\} &       3 (4) \{32\} &      5 (6) \{32\} &          \{4\} &       8 (9) \{56\} &    3/11 (27\%) \\
17  &   &	&   Y2 &      4 (6) \{333\} &       3 (3) \{24\} &      2 (2) \{54\} &         \{90\} &      4 (4) \{130\} &      0/5 (0\%) \\
\hline
4   &   \multirow{2}{*}{FI0009005805} & \multirow{2}{*}{1997-06-09} &   Y1 &    28 (44) \{240\} &     16 (16) \{35\} &    22 (19) \{39\} &       8 (7) \{14\} &     19 (22) \{53\} &    4/39 (10\%) \\
5   &   &	&   Y2 &     9 (13) \{356\} &       3 (3) \{29\} &      4 (7) \{62\} &       5 (7) \{93\} &      5 (9) \{160\} &    2/12 (17\%) \\
\hline
18  &   \multirow{2}{*}{FI0009006589} & \multirow{2}{*}{1997-06-17} &   Y1 &     9 (14) \{245\} &       7 (6) \{39\} &      6 (7) \{37\} &       4 (4) \{14\} &      6 (11) \{51\} &    2/11 (18\%) \\
19  &   &	&   Y2 &      1 (1) \{377\} &         \{29\} &        \{62\} &         \{91\} &        \{162\} &      0/1 (0\%) \\
\hline
20  &   \multirow{2}{*}{FI0009006621} & \multirow{2}{*}{1997-11-25} &   Y1 &    60 (81) \{273\} &     23 (19) \{27\} &    24 (20) \{46\} &     31 (24) \{43\} &     44 (39) \{62\} &     3/66 (5\%) \\
21  &   &	&  Y2 &   59 (123) \{547\} &     26 (20) \{27\} &    24 (21) \{46\} &     50 (55) \{90\} &   53 (101) \{236\} &     2/63 (3\%) \\
\hline
24  &   \multirow{2}{*}{FI0009006738} & \multirow{2}{*}{1997-11-26} &   Y1 &    30 (46) \{275\} &     11 (10) \{28\} &    18 (16) \{48\} &     16 (15) \{45\} &     20 (21) \{62\} &     3/38 (8\%) \\
25  &   &	&  Y2 &   54 (107) \{557\} &     12 (12) \{27\} &    14 (14) \{47\} &     43 (48) \{97\} &    57 (86) \{241\} &     5/71 (7\%) \\
\hline
22  &   \multirow{2}{*}{FI0009006696} & \multirow{2}{*}{1997-12-02} &   Y1 &      4 (5) \{274\} &       1 (1) \{28\} &      1 (1) \{48\} &       2 (3) \{46\} &       4 (5) \{65\} &      0/5 (0\%) \\
23  &   &	&   Y2 &      5 (8) \{602\} &       3 (3) \{25\} &      4 (4) \{50\} &       3 (3) \{88\} &      5 (9) \{210\} &      0/5 (0\%) \\
\hline
26  &   \multirow{2}{*}{FI0009006761} & \multirow{2}{*}{1997-12-09} &   Y1 &     8 (14) \{291\} &       6 (8) \{26\} &     8 (10) \{49\} &       7 (9) \{45\} &       6 (8) \{68\} &      0/8 (0\%) \\
27  &   &	&   Y2 &     7 (15) \{606\} &       6 (5) \{25\} &      4 (4) \{41\} &      6 (10) \{90\} &     7 (18) \{229\} &      0/7 (0\%) \\
\hline
32  &   \multirow{2}{*}{FI0009007017} & \multirow{2}{*}{1998-04-01} &   Y1 &        \{316\} &         \{28\} &        \{62\} &         \{81\} &        \{117\} &      0  \\
33  &   &	&   Y2 &      3 (4) \{765\} &       2 (2) \{32\} &        \{44\} &       1 (1) \{87\} &      3 (4) \{232\} &     1/4 (25\%) \\
\hline
34  &   \multirow{2}{*}{FI0009007025} & \multirow{2}{*}{1998-04-01} &   Y1 &      7 (9) \{316\} &       6 (6) \{28\} &      5 (6) \{62\} &       6 (6) \{81\} &      6 (7) \{117\} &     2/9 (22\%) \\
35  &   &	&   Y2 &    15 (24) \{765\} &     10 (11) \{32\} &     10 (9) \{44\} &     10 (16) \{87\} &    14 (22) \{232\} &    3/20 (15\%) \\
\hline
36  &   \multirow{2}{*}{FI0009007066} & \multirow{2}{*}{1998-04-30} &   Y1 &      6 (9) \{326\} &       5 (7) \{29\} &      4 (5) \{56\} &       5 (8) \{87\} &     6 (11) \{142\} &      0/8 (0\%) \\
37  &   &	&   Y2 &      3 (2) \{836\} &       2 (1) \{35\} &      2 (3) \{42\} &      2 (2) \{102\} &      2 (2) \{223\} &      0/3 (0\%) \\
\hline
28  &   \multirow{2}{*}{FI0009006829} & \multirow{2}{*}{1998-06-01} &   Y1 &      4 (5) \{341\} &       3 (3) \{30\} &      2 (2) \{60\} &       2 (2) \{89\} &      1 (1) \{153\} &      0/5 (0\%) \\
29  &   &	&   Y2 &     9 (14) \{906\} &       4 (4) \{38\} &     7 (12) \{46\} &       5 (6) \{96\} &      7 (7) \{210\} &    5/15 (33\%) \\
\hline
40  &   \multirow{2}{*}{FI0009007215} & \multirow{2}{*}{1998-08-03} &   Y1 &     6 (12) \{437\} &       1 (2) \{30\} &      5 (4) \{57\} &       4 (6) \{93\} &     6 (11) \{179\} &    3/10 (30\%) \\
41  &   &	&   Y2 &    15 (24) \{615\} &       6 (6) \{39\} &      5 (6) \{45\} &    13 (14) \{103\} &    11 (12) \{185\} &    6/25 (24\%) \\
\hline
42  &   \multirow{2}{*}{FI0009007264} & \multirow{2}{*}{1998-09-15} &   Y1 &   94 (134) \{469\} &     16 (13) \{25\} &    15 (15) \{48\} &     53 (45) \{94\} &    75 (81) \{204\} &    9/113 (8\%) \\
43  &   &	&   Y2 &  376 (361) \{587\} &     37 (31) \{41\} &    32 (23) \{41\} &  163 (105) \{125\} &  237 (144) \{200\} &  64/475 (13\%) \\
\hline
46  &   \multirow{2}{*}{FI0009007371} & \multirow{2}{*}{1998-11-17} &   Y1 &  248 (350) \{515\} &     41 (24) \{26\} &    49 (35) \{45\} &    148 (83) \{90\} &  206 (187) \{233\} &   13/272 (5\%) \\
47  &   &	&   Y2 &  683 (550) \{645\} &     47 (31) \{36\} &    49 (39) \{48\} &  252 (108) \{112\} &  300 (156) \{183\} &  99/818 (12\%) \\
\hline
44  &   \multirow{2}{*}{FI0009007355} & \multirow{2}{*}{1998-12-04} &   Y1 &      1 (1) \{612\} &         \{26\} &        \{47\} &         \{91\} &      1 (1) \{226\} &      0/1 (0\%) \\
45  &   &	&   Y2 &        \{611\} &         \{33\} &        \{44\} &        \{115\} &        \{175\} &      0  \\
\hline
38  &   \multirow{2}{*}{FI0009007132} & \multirow{2}{*}{1998-12-18} &   Y1 &   91 (153) \{629\} &     26 (16) \{24\} &    32 (23) \{40\} &     67 (60) \{87\} &   85 (125) \{250\} &    9/111 (8\%) \\
39  &   &	&   Y2 &    35 (45) \{523\} &     18 (19) \{33\} &    29 (25) \{49\} &    31 (34) \{112\} &    38 (58) \{166\} &     4/54 (7\%) \\
\hline
52  &   \multirow{2}{*}{FI0009007629} & \multirow{2}{*}{1999-03-01} &   Y1 &      2 (3) \{689\} &       1 (1) \{33\} &        \{41\} &         \{81\} &      1 (1) \{261\} &      0/2 (0\%) \\
53  &   &	&   Y2 &      6 (7) \{553\} &       3 (3) \{34\} &      4 (4) \{56\} &       8 (9) \{97\} &      4 (4) \{130\} &    3/12 (25\%) \\
\hline
126 &   \multirow{2}{*}{FI0009801286} & \multirow{2}{*}{1999-03-15} &   Y1 &    14 (23) \{731\} &       6 (5) \{33\} &      5 (5) \{41\} &     10 (13) \{81\} &    15 (28) \{262\} &     0/15 (0\%) \\
127 &   &	&   Y2 &      1 (1) \{580\} &       1 (1) \{32\} &        \{54\} &       1 (1) \{96\} &      1 (1) \{124\} &      0/1 (0\%) \\
\hline
50  &   \multirow{2}{*}{FI0009007553} & \multirow{2}{*}{1999-03-23} &   Y1 &   65 (108) \{745\} &     28 (19) \{35\} &    20 (17) \{48\} &     44 (44) \{83\} &    61 (77) \{242\} &    8/81 (10\%) \\
51  &   &	&   Y2 &    65 (70) \{551\} &     12 (10) \{36\} &    14 (13) \{52\} &    56 (48) \{100\} &    42 (26) \{128\} &   11/99 (11\%) \\
\hline
58  &   \multirow{2}{*}{FI0009007728} & \multirow{2}{*}{1999-04-06} &   Y1 &    36 (78) \{811\} &     16 (20) \{33\} &     8 (10) \{46\} &     27 (32) \{92\} &    36 (59) \{230\} &    5/44 (11\%) \\
59  &   &	&   Y2 &    21 (22) \{530\} &     10 (11) \{32\} &    10 (10) \{53\} &     18 (22) \{90\} &    11 (11) \{125\} &    5/31 (16\%) \\
\hline
48  &   \multirow{2}{*}{FI0009007546} & \multirow{2}{*}{1999-04-19} &   Y1 &      1 (1) \{850\} &       1 (1) \{35\} &      1 (1) \{46\} &        \{100\} &      1 (1) \{225\} &      0/1 (0\%) \\
49  &   &	&   Y2 &        \{447\} &         \{29\} &        \{53\} &         \{84\} &        \{124\} &      0  \\
\hline
56  &   \multirow{2}{*}{FI0009007694} & \multirow{2}{*}{1999-05-03} &   Y1 &    23 (43) \{855\} &     11 (13) \{37\} &      4 (6) \{44\} &     18 (28) \{99\} &    19 (29) \{228\} &    4/27 (15\%) \\
57  &   &	&   Y2 &      1 (2) \{454\} &       1 (2) \{34\} &        \{51\} &       2 (3) \{93\} &        \{131\} &      0/2 (0\%) \\
\hline
54  &   \multirow{2}{*}{FI0009007686} & \multirow{2}{*}{1999-05-03} &   Y1 &      3 (7) \{855\} &         \{37\} &        \{44\} &       3 (3) \{99\} &      4 (7) \{228\} &     1/5 (20\%) \\
55  &   &	&   Y2 &        \{454\} &         \{34\} &        \{51\} &         \{93\} &        \{131\} &    1/1 (100\%) \\
\hline
30  &   \multirow{2}{*}{FI0009006886} & \multirow{2}{*}{1999-06-08} &   Y1 &      6 (6) \{898\} &       2 (2) \{33\} &      3 (3) \{47\} &       5 (6) \{97\} &     8 (10) \{210\} &    2/11 (18\%) \\
31  &   &	&   Y2 &      2 (2) \{433\} &       1 (1) \{34\} &      1 (1) \{53\} &       1 (1) \{90\} &      1 (1) \{130\} &      0/2 (0\%) \\
\hline
60  &   \multirow{2}{*}{FI0009007819} & \multirow{2}{*}{1999-06-28} &   Y1 &  122 (277) \{859\} &     38 (23) \{36\} &    29 (28) \{47\} &     86 (63) \{92\} &   98 (115) \{203\} &    5/135 (4\%) \\
61  &   &	&   Y2 &  101 (124) \{513\} &     31 (20) \{34\} &    35 (33) \{51\} &     81 (67) \{98\} &    63 (58) \{123\} &    6/123 (5\%) \\
\hline
62  &   \multirow{2}{*}{FI0009007835} & \multirow{2}{*}{1999-07-01} &   Y1 &    35 (64) \{791\} &     22 (22) \{34\} &    16 (18) \{48\} &     30 (38) \{95\} &    30 (44) \{186\} &    4/41 (10\%) \\
63  &   &	&   Y2 &    30 (42) \{596\} &     19 (17) \{34\} &    16 (16) \{51\} &    23 (31) \{103\} &    18 (27) \{124\} &     1/34 (3\%) \\
\hline
64  &   \multirow{2}{*}{FI0009007884} & \multirow{2}{*}{1999-07-01} &   Y1 &  104 (171) \{791\} &     31 (23) \{34\} &    27 (26) \{48\} &     74 (57) \{95\} &    82 (75) \{186\} &  13/136 (10\%) \\
65  &   &	&   Y2 &   82 (145) \{596\} &     32 (24) \{34\} &    36 (29) \{51\} &    73 (73) \{103\} &    57 (61) \{124\} &    6/100 (6\%) \\
\hline
68  &   \multirow{2}{*}{FI0009008080} & \multirow{2}{*}{1999-10-01} &   Y1 &     9 (16) \{597\} &       3 (4) \{41\} &      2 (1) \{42\} &      6 (7) \{120\} &      7 (8) \{197\} &    2/11 (18\%) \\
69  &   &	&   Y2 &     9 (15) \{775\} &       5 (7) \{34\} &      5 (6) \{53\} &      7 (10) \{89\} &      7 (7) \{111\} &      0/9 (0\%) \\
\hline
66  &   \multirow{2}{*}{FI0009007918} & \multirow{2}{*}{1999-10-27} &   Y1 &   81 (167) \{631\} &     20 (21) \{41\} &    16 (16) \{47\} &    58 (64) \{120\} &    58 (64) \{192\} &     2/87 (2\%) \\
67  &   &	&   Y2 &   67 (150) \{763\} &     21 (22) \{37\} &    29 (21) \{51\} &     51 (61) \{91\} &    33 (44) \{101\} &     6/79 (8\%) \\
\hline
128 &   \multirow{2}{*}{FI0009801310} & \multirow{2}{*}{1999-11-09} &   Y1 &  136 (149) \{630\} &     22 (18) \{38\} &    20 (16) \{50\} &    85 (58) \{113\} &    69 (67) \{195\} &   14/169 (8\%) \\
129 &   &	&   Y2 &  183 (318) \{765\} &     34 (21) \{38\} &    35 (30) \{45\} &     90 (67) \{86\} &    93 (70) \{108\} &  22/218 (10\%) \\
\hline
70  &   \multirow{2}{*}{FI0009008221} & \multirow{2}{*}{1999-12-13} &   Y1 &  280 (292) \{536\} &     37 (23) \{34\} &    39 (29) \{46\} &   151 (91) \{112\} &  146 (108) \{171\} &   32/337 (9\%) \\
71  &   &	&   Y2 &  215 (356) \{843\} &     50 (27) \{39\} &    52 (30) \{42\} &    115 (74) \{89\} &    78 (60) \{104\} &   22/252 (9\%) \\
\hline
132 &   \multirow{2}{*}{FI0009902530} & \multirow{2}{*}{2000-01-31} &   Y1 &    46 (61) \{489\} &     17 (22) \{37\} &    21 (21) \{55\} &    39 (49) \{115\} &    29 (46) \{158\} &     5/62 (8\%) \\
133 &   &	&   Y2 &   55 (118) \{915\} &     34 (29) \{49\} &    37 (31) \{46\} &     40 (47) \{90\} &     20 (19) \{83\} &     2/65 (3\%) \\
\hline
76  &   \multirow{2}{*}{FI0009008924} & \multirow{2}{*}{2000-05-24} &   Y1 &    10 (15) \{439\} &       4 (4) \{33\} &      3 (3) \{58\} &       6 (7) \{86\} &      7 (8) \{130\} &     0/12 (0\%) \\
77  &   &	&   Y2 &      1 (1) \{856\} &         \{61\} &        \{47\} &         \{80\} &         \{70\} &     1/2 (50\%) \\
\hline
74  &   \multirow{2}{*}{FI0009008833} & \multirow{2}{*}{2000-05-24} &   Y1 &        \{439\} &         \{33\} &      1 (1) \{58\} &       1 (2) \{86\} &        \{130\} &     1/2 (50\%) \\
75  &   &	&   Y2 &      2 (2) \{856\} &       2 (2) \{61\} &      2 (3) \{47\} &       2 (2) \{80\} &         \{70\} &      0/3 (0\%) \\
\hline
80  &   \multirow{2}{*}{FI0009009146} & \multirow{2}{*}{2000-07-04} &   Y1 &    18 (28) \{586\} &       4 (4) \{35\} &      7 (8) \{54\} &     15 (19) \{98\} &    13 (15) \{123\} &    5/28 (18\%) \\
81  &   &	&   Y2 &      1 (1) \{849\} &       1 (1) \{61\} &      1 (1) \{46\} &       1 (1) \{78\} &         \{66\} &      0/1 (0\%) \\
\hline
78  &   \multirow{2}{*}{FI0009009054} & \multirow{2}{*}{2000-07-05} &   Y1 &      5 (8) \{600\} &       2 (4) \{35\} &      1 (1) \{54\} &      4 (6) \{102\} &      3 (3) \{128\} &     1/8 (12\%) \\
79  &   &	&   Y2 &      3 (4) \{849\} &       3 (4) \{60\} &      3 (4) \{47\} &       2 (5) \{78\} &       4 (4) \{69\} &      0/8 (0\%) \\
\hline
86  &   \multirow{2}{*}{FI0009009633} & \multirow{2}{*}{2000-11-01} &   Y1 &    12 (16) \{762\} &       3 (4) \{37\} &      3 (4) \{47\} &       6 (6) \{92\} &      7 (6) \{107\} &    5/19 (26\%) \\
87  &   &	&   Y2 &      7 (9) \{548\} &       2 (2) \{64\} &      1 (1) \{50\} &       4 (4) \{93\} &       2 (2) \{53\} &    5/14 (36\%) \\
\hline
84  &   \multirow{2}{*}{FI0009009567} & \multirow{2}{*}{2000-12-19} &   Y1 &      8 (9) \{818\} &       5 (5) \{40\} &      4 (2) \{47\} &       8 (8) \{93\} &      1 (1) \{101\} &    3/12 (25\%) \\
85  &   &	&   Y2 &      2 (4) \{468\} &       4 (4) \{58\} &      3 (3) \{59\} &      5 (5) \{101\} &       2 (2) \{52\} &      0/8 (0\%) \\
\hline
72  &   \multirow{2}{*}{FI0009008270} & \multirow{2}{*}{2000-12-22} &   Y1 &    45 (68) \{858\} &     16 (12) \{39\} &    17 (11) \{48\} &     23 (26) \{91\} &     17 (18) \{91\} &     5/53 (9\%) \\
73  &   &	&   Y2 &      4 (4) \{477\} &       5 (5) \{59\} &      2 (3) \{58\} &       6 (5) \{95\} &       3 (3) \{52\} &    2/12 (17\%) \\
\hline
88  &   \multirow{2}{*}{FI0009009674} & \multirow{2}{*}{2001-01-30} &   Y1 &    18 (36) \{900\} &     19 (24) \{47\} &    19 (25) \{52\} &     19 (31) \{91\} &     11 (11) \{81\} &     1/24 (4\%) \\
89  &   &	&   Y2 &    11 (24) \{499\} &     11 (19) \{58\} &    11 (22) \{52\} &    12 (35) \{105\} &       4 (6) \{41\} &     1/13 (8\%) \\
\hline
82  &   \multirow{2}{*}{FI0009009377} & \multirow{2}{*}{2001-04-02} &   Y1 &      2 (2) \{795\} &       2 (2) \{54\} &      1 (1) \{49\} &       2 (2) \{88\} &      1 (2) \{102\} &     2/5 (40\%) \\
83  &   &	&   Y2 &      1 (1) \{526\} &         \{42\} &      2 (2) \{53\} &      1 (1) \{118\} &         \{14\} &     3/6 (50\%) \\
\hline
90  &   \multirow{2}{*}{FI0009010219} & \multirow{2}{*}{2001-04-02} &   Y1 &     5 (11) \{795\} &       4 (4) \{54\} &      4 (5) \{49\} &       5 (5) \{88\} &      4 (6) \{102\} &     1/8 (12\%) \\
91  &   &	&   Y2 &      2 (2) \{526\} &       1 (1) \{42\} &        \{53\} &      1 (1) \{118\} &         \{14\} &     2/5 (40\%) \\
\hline
96  &   \multirow{2}{*}{FI0009010862} & \multirow{2}{*}{2001-10-01} &   Y1 &      1 (1) \{639\} &       5 (6) \{62\} &      4 (5) \{55\} &      3 (4) \{101\} &       1 (1) \{51\} &      0/5 (0\%) \\
97  &   &	&   Y2 &        \{164\} &       3 (3) \{39\} &      5 (8) \{67\} &       7 (8) \{94\} &         \{14\} &    2/10 (20\%) \\
\hline
94  &   \multirow{2}{*}{FI0009010854} & \multirow{2}{*}{2001-10-01} &   Y1 &      2 (2) \{639\} &       2 (3) \{62\} &      2 (2) \{55\} &      2 (3) \{101\} &         \{51\} &      0/2 (0\%) \\
95  &   &	&   Y2 &      3 (3) \{164\} &       5 (5) \{39\} &    12 (15) \{67\} &     13 (20) \{94\} &         \{14\} &     0/14 (0\%) \\
\hline
136 &  \multirow{2}{*}{SE0000667925} & \multirow{2}{*}{2002-12-09} &   Y1 &    42 (44) \{248\} &     18 (18) \{47\} &    29 (27) \{57\} &     57 (51) \{91\} &     20 (18) \{35\} &  33/120 (28\%) \\
137 &   &	&   Y2 &   96 (125) \{327\} &     52 (45) \{58\} &    60 (49) \{70\} &     66 (63) \{86\} &     45 (37) \{73\} &   11/129 (9\%) \\
\hline
134 &  \multirow{2}{*}{SE0000110165} & \multirow{2}{*}{2003-09-04} &   Y1 &      1 (1) \{339\} &         \{44\} &      2 (3) \{66\} &       1 (1) \{93\} &       1 (1) \{77\} &     2/4 (50\%) \\
135 &   &	&   Y2 &        \{173\} &         \{56\} &       \{105\} &        \{100\} &         \{73\} &      0  \\
\hline
100 &   \multirow{2}{*}{FI0009012843} & \multirow{2}{*}{2004-10-18} &   Y1 &    23 (41) \{187\} &     37 (42) \{71\} &   34 (44) \{107\} &     36 (53) \{99\} &     20 (27) \{71\} &    7/54 (13\%) \\
101 &   &	&  Y2 &    41 (65) \{438\} &    36 (42) \{159\} &   42 (54) \{133\} &    31 (33) \{113\} &       6 (5) \{25\} &     6/64 (9\%) \\
\hline
102 &   \multirow{2}{*}{FI0009013296} & \multirow{2}{*}{2005-04-21} &   Y1 &  153 (138) \{211\} &   144 (87) \{102\} &    76 (60) \{95\} &    109 (65) \{81\} &     31 (28) \{63\} &  36/262 (14\%) \\
103 &   &	&   Y2 &  237 (207) \{273\} &  165 (134) \{185\} &  106 (76) \{107\} &   148 (89) \{110\} &     34 (30) \{60\} &  41/336 (12\%) \\
\hline
104 &   \multirow{2}{*}{FI0009013312} & \multirow{2}{*}{2005-06-01} &   Y1 &    19 (28) \{323\} &    13 (14) \{129\} &   18 (17) \{107\} &     15 (13) \{82\} &      9 (12) \{68\} &     1/26 (4\%) \\
105 &   &	&   Y2 &      6 (9) \{266\} &      6 (7) \{171\} &      7 (8) \{95\} &      8 (10) \{94\} &       5 (4) \{50\} &     1/13 (8\%) \\
\hline
106 &   \multirow{2}{*}{FI0009013403} & \multirow{2}{*}{2005-06-01} &   Y1 &    75 (85) \{323\} &    81 (75) \{129\} &   39 (37) \{107\} &     56 (45) \{82\} &     22 (15) \{68\} &   10/112 (9\%) \\
107 &   &	&   Y2 &    64 (64) \{266\} &    66 (80) \{171\} &    34 (31) \{95\} &     55 (49) \{94\} &       7 (8) \{50\} &   10/92 (11\%) \\
\hline
108 &   \multirow{2}{*}{FI0009013429} & \multirow{2}{*}{2005-06-01} &   Y1 &   88 (105) \{323\} &    85 (77) \{129\} &   51 (44) \{107\} &     71 (54) \{82\} &     21 (16) \{68\} &  18/133 (14\%) \\
109 &   &	&   Y2 &    61 (73) \{266\} &    64 (69) \{171\} &    25 (26) \{95\} &     48 (39) \{94\} &      9 (10) \{50\} &    9/89 (10\%) \\
\hline
92  &   \multirow{2}{*}{FI0009010391} & \multirow{2}{*}{2006-03-17} &   Y1 &    21 (25) \{243\} &    16 (17) \{182\} &   22 (26) \{108\} &    20 (21) \{119\} &     12 (14) \{68\} &    4/37 (11\%) \\
93  &   &	&   Y2 &    22 (26) \{370\} &    17 (19) \{121\} &   28 (37) \{118\} &    26 (34) \{101\} &       9 (8) \{36\} &    6/50 (12\%) \\
\hline
112 &   \multirow{2}{*}{FI0009013924} & \multirow{2}{*}{2006-03-17} &   Y1 &    19 (22) \{243\} &    13 (12) \{182\} &   17 (22) \{108\} &    16 (17) \{119\} &       5 (5) \{68\} &    3/29 (10\%) \\
113 &   &	&   Y2 &      4 (4) \{370\} &      4 (4) \{121\} &     2 (2) \{118\} &      4 (3) \{101\} &       1 (1) \{36\} &    2/11 (18\%) \\
\hline
110 &   \multirow{2}{*}{FI0009013593} & \multirow{2}{*}{2006-04-21} &   Y1 &    10 (14) \{268\} &      6 (7) \{184\} &     5 (9) \{102\} &    10 (14) \{112\} &       2 (2) \{60\} &    2/17 (12\%) \\
111 &   &	&   Y2 &        \{337\} &        \{127\} &       \{124\} &         \{90\} &         \{32\} &      0  \\
\hline
116 &   \multirow{2}{*}{FI0009014351} & \multirow{2}{*}{2006-07-03} &   Y1 &    28 (32) \{263\} &    21 (24) \{163\} &    25 (27) \{93\} &     26 (25) \{90\} &       6 (6) \{50\} &   16/56 (29\%) \\
117 &   &	&   Y2 &      3 (4) \{336\} &      4 (4) \{147\} &     5 (5) \{142\} &       4 (4) \{94\} &         \{34\} &      0/6 (0\%) \\
\hline
114 &   \multirow{2}{*}{FI0009014344} & \multirow{2}{*}{2006-07-03} &   Y1 &        \{263\} &        \{163\} &        \{93\} &         \{90\} &         \{50\} &    3/3 (100\%) \\
115 &   &	&   Y2 &        \{336\} &        \{147\} &       \{142\} &         \{94\} &         \{34\} &      0  \\
\hline
98  &   \multirow{2}{*}{FI0009012413} & \multirow{2}{*}{2007-04-10} &   Y1 &      4 (4) \{360\} &      2 (3) \{122\} &     9 (9) \{122\} &       6 (4) \{95\} &       3 (3) \{38\} &    8/22 (36\%) \\
99  &   &	&   Y2 &      6 (7) \{315\} &    11 (13) \{143\} &    9 (16) \{161\} &     10 (13) \{62\} &          \{7\} &    6/22 (27\%) \\
\hline
118 &   \multirow{2}{*}{FI0009015309} & \multirow{2}{*}{2007-06-15} &   Y1 &    32 (37) \{329\} &    26 (26) \{139\} &   31 (35) \{136\} &     34 (32) \{92\} &       6 (5) \{37\} &   12/64 (19\%) \\
119 &   &	&   Y2 &    19 (26) \{380\} &    23 (30) \{211\} &   16 (24) \{198\} &     14 (17) \{67\} &       2 (2) \{16\} &     2/29 (7\%) \\
\midrule
 &   &  &    &     &     &    &     \multirow{2}{*}{\textbf{Median}} &    \textbf{Y1}    & 11\% \\
  &   &  &    &     &     &    &    &    \textbf{Y2}    & 9\% \\
  \midrule
 &   &  &    &     &     &    &   \multirow{2}{*}{\textbf{Average}} &    \textbf{Y1}    & 16\% \\
  &   &  &    &     &     &    &    &    \textbf{Y2}    & 13\% \\
\bottomrule
\end{longtabu}
}

\section*{Conclusions}
In the current paper, we analysed investor interactions and behaviours using a unique dataset of all Finnish investors’ transactions in the HSE.
Our selected set of 69 securities is aligned to an IPO event, which occurs when a company first starts publicly trading its securities. 
We performed an analysis for multiple securities on an individual investor account level by constructing the networks from the statistically validated trading co-occurrences.
Our main focus was on the newly emerging market networks and their common and persistent market-driven structures with the other mature and new stocks.

Applying a community detection algorithm, we found statistically similar investor clusters with synchronised trading strategies that were forming repeatedly over several years and for multiple securities.
We detected statistically robust clusters between the first and second year after an IPO. 
We also found clusters that could be found within other securities.
By investigating cluster attribute over- and underexpression, we find a highly persistent institutional investor cluster. This finding provides further evidence about institutional herding.
Comparing the findings with the clusters on mature securities, we observe that the majority of clusters can also be observed with a mature security.

Our results show that some synchronised trading strategies in financial markets span across multiple stocks, are persistent over time and occur with both newly issued and mature stocks. 
However, this analysis applies to the HSE only and does not generalise to all markets. 
Further research should check if this phenomenon also exists in other stock exchanges with a larger amount of IPOs; however, to the best of our knowledge, these investor-level data are not available, for example, from the U.S. markets.

Traditional financial research  assumes that investors are rational and hold optimal portfolios. 
However, actual investors have information, intellectual and computational limitations, and they satisfice\footnote{The term \textit{satisfice} refers to making optimal decisions under the limited resources. It was first defined in \citep{simon1947administrative}.} when making decisions.
The systematic reoccurrence of the clusters gives a notion of possible stronger information connections that the investors share. 
For example, they may be consistently following the same public information sources or have mutual private information channels. 
However, with the current research, we do not try to explain the direction or the publicity of the information transfer.
On the other hand, according to \citet{ozsoylev2013investor}, investor networks can be considered proxies of information networks if they are fairly stable over time. 
In light of this argument, the persistent and security-wide investor clusters can represent the mutual information channels that exist for both new IPO securities and mature stocks (e.g., Nokia).

\section*{Data availability}
The dataset analysed in the current study is not publicly available and cannot be distributed by the authors because it is a proprietary database of Euroclear Finland. The database can be accessed for research purposes under the nondisclosure agreement by asking permission from Euroclear Finland.

\section*{Acknowledgements}
M.B. is grateful for the grants received from the Finnish Foundation for Technology Promotion and the Finnish Foundation for Share Promotion. 
K.B. received funding from the EU Research and Innovation Programme Horizon 2020 under grant agreement No. 675044 (BigDataFinance) and from the doctoral school of Tampere University.
F.L. and D.P. acknowledge partial support from the European Community H2020 Program under the scheme INFRAIA-1-2014-2015: Research Infrastructures, grant agreement No. 654024 SoBigData: Social Mining and Big Data Ecosystem (\href{http://www.sobigdata.eu}{http://www.sobigdata.eu}). 
The funders had no role in study design, data collection and analysis, decision to publish or preparation of the manuscript. 

\nolinenumbers
\pagebreak

\bibliographystyle{apalike}
\bibliography{CommunityOverlaps}

\begin{figure}[H]
    \centering
    \includegraphics[width=0.95\textwidth]{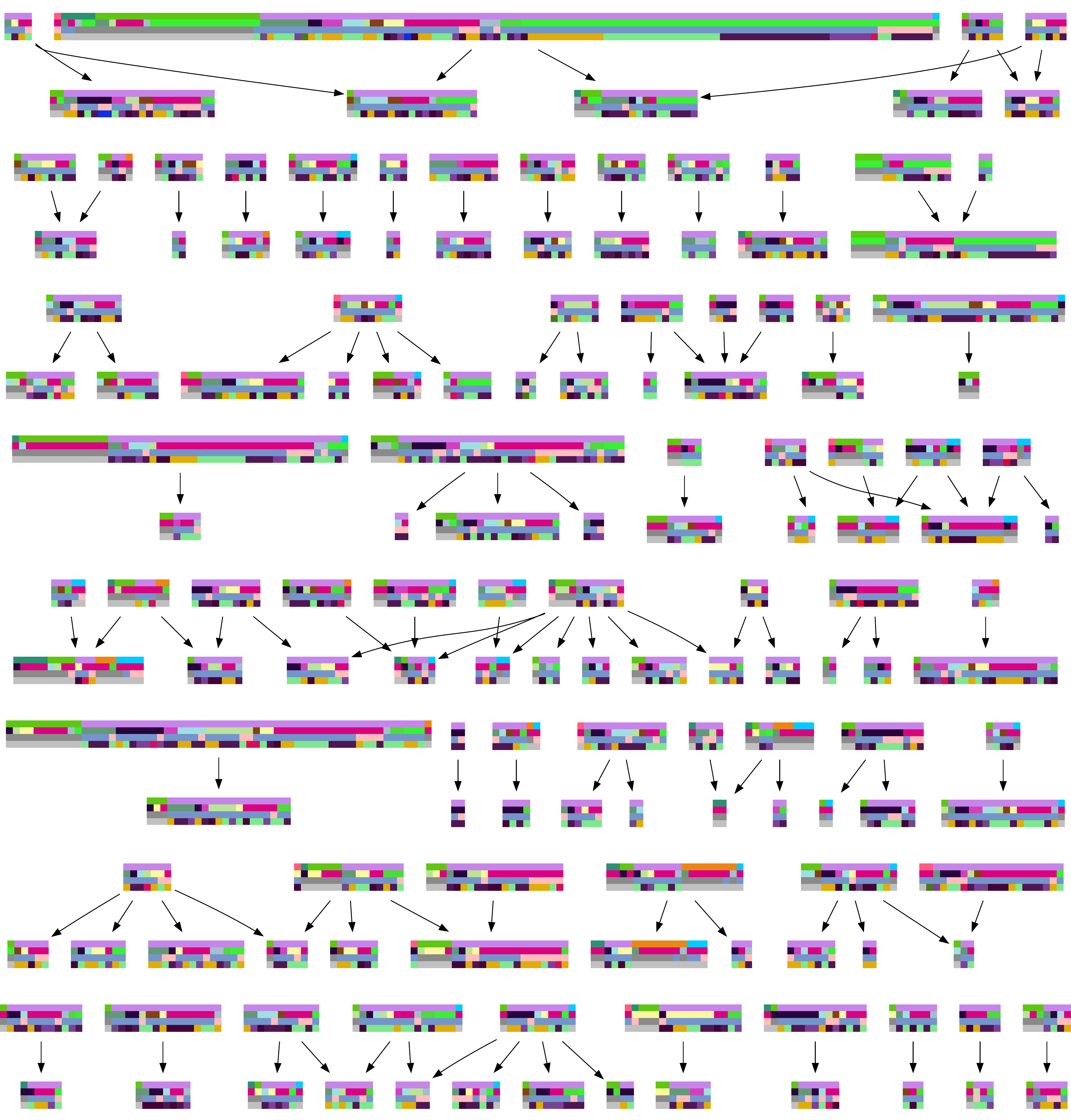}
  \caption {
  {\bf Cluster evolution for networks with the FDR validation.}
  \label{fig:netw_infomap_evol1}  Partial results for ISIN FI0009013296.  
 A cluster is represented by a rectangle. 
 Cluster evolution is represented by the rectangles connected by downward arrows.
 The top rectangle is the cluster in the first year after the IPO, and the bottom rectangle is the cluster in the second year after an IPO in the same network. 
\textbf{Sector code:}
\protect\tikz{\protect\draw [black, thin, fill=cc686e9] (4pt,4pt) rectangle (0.4,0.4) ;} - Households, 
\protect\tikz{\protect\draw [black, thin, fill=c5fc613] (4pt,4pt) rectangle (0.4,0.4);} - Non-financial, 
\protect\tikz{\protect\draw [black, thin, fill=c00caff] (4pt,4pt) rectangle (0.4,0.4);} - Financial-Insurance, 
\protect\tikz{\protect\draw [black, thin, fill=cea8615] (4pt,4pt) rectangle (0.4,0.4);} - General-Government, 
\protect\tikz{\protect\draw [black, thin, fill=c2e9072] (4pt,4pt) rectangle (0.4,0.4);} - Non-Profit, 
\protect\tikz{\protect\draw [black, thin, fill=cff5c81] (4pt,4pt) rectangle (0.4,0.4);} - Rest-World. \\
\textbf{Geographic location:}
\protect\tikz{\protect\draw [black, thin, fill=cda027f] (4pt,4pt) rectangle (0.4,0.4) ;} - Helsinki, 
\protect\tikz{\protect\draw [black, thin, fill=c29033c] (4pt,4pt) rectangle (0.4,0.4);} - South-West, 
\protect\tikz{\protect\draw [black, thin, fill=c639a76] (4pt,4pt) rectangle (0.4,0.4);} - Western-Tavastia, 
\protect\tikz{\protect\draw [black, thin, fill=c36f230] (4pt,4pt) rectangle (0.4,0.4);} - Central-Finland, 
\protect\tikz{\protect\draw [black, thin, fill=cf9f99d] (4pt,4pt) rectangle (0.4,0.4);} - Northern-Finland, 
\protect\tikz{\protect\draw [black, thin, fill=cbae396] (4pt,4pt) rectangle (0.4,0.4);} - Ostrobothnia,
\protect\tikz{\protect\draw [black, thin, fill=c9ee0e0] (4pt,4pt) rectangle (0.4,0.4) ;} - Rest-Uusimaa, 
\protect\tikz{\protect\draw [black, thin, fill=ca5bdcc] (4pt,4pt) rectangle (0.4,0.4);} - Eastern-Tavastia, 
\protect\tikz{\protect\draw [black, thin, fill=c4cdc38] (4pt,4pt) rectangle (0.4,0.4) ;} - Eastern-Finland, 
\protect\tikz{\protect\draw [black, thin, fill=cd340bd] (4pt,4pt) rectangle (0.4,0.4);} - South-East, 
\protect\tikz{\protect\draw [black, thin, fill=c874010] (4pt,4pt) rectangle (0.4,0.4);} - Northern-Savonia. \\
\textbf{Gender:}
\protect\tikz{\protect\draw [black, thin, fill=c7495cb] (4pt,4pt) rectangle (0.4,0.4) ;} - Male, 
\protect\tikz{\protect\draw [black, thin, fill=cffbcbc] (4pt,4pt) rectangle (0.4,0.4);} - Female, 
\protect\tikz{\protect\draw [black, thin, fill=c8a8a8a] (4pt,4pt) rectangle (0.4,0.4);} - No-Gender. \\
\textbf{Decade:}
\protect\tikz{\protect\draw [black, thin, fill=cc0c0c0] (4pt,4pt) rectangle (0.4,0.4) ;} - No-Age, 
\protect\tikz{\protect\draw [black, thin, fill=c4b771a] (4pt,4pt) rectangle (0.4,0.4);} - 1910, 
\protect\tikz{\protect\draw [black, thin, fill=cda0c5a] (4pt,4pt) rectangle (0.4,0.4);} - 1920,
\protect\tikz{\protect\draw [black, thin, fill=c7e3f9c] (4pt,4pt) rectangle (0.4,0.4) ;} - 1930, 
\protect\tikz{\protect\draw [black, thin, fill=c501653] (4pt,4pt) rectangle (0.4,0.4);} - 1940, 
\protect\tikz{\protect\draw [black, thin, fill=c7de88c] (4pt,4pt) rectangle (0.4,0.4);} - 1950,
\protect\tikz{\protect\draw [black, thin, fill=ce1ac06] (4pt,4pt) rectangle (0.4,0.4) ;} - 1960, 
\protect\tikz{\protect\draw [black, thin, fill=c430337] (4pt,4pt) rectangle (0.4,0.4);} - 1970, 
\protect\tikz{\protect\draw [black, thin, fill=c6e4b83] (4pt,4pt) rectangle (0.4,0.4);} - 1980,
\protect\tikz{\protect\draw [black, thin, fill=ce40030] (4pt,4pt) rectangle (0.4,0.4);} - 1990, 
\protect\tikz{\protect\draw [black, thin, fill=c102ee2] (4pt,4pt) rectangle (0.4,0.4);} - 2000.}
\end{figure}

\begin{figure}[!ht]
    \centering  
    \captionsetup[subfigure]{justification=centering}
    
        \begin{multicols}{2}
    \begin{subfigure}{1\textwidth}
      \includegraphics[width=1\textwidth]{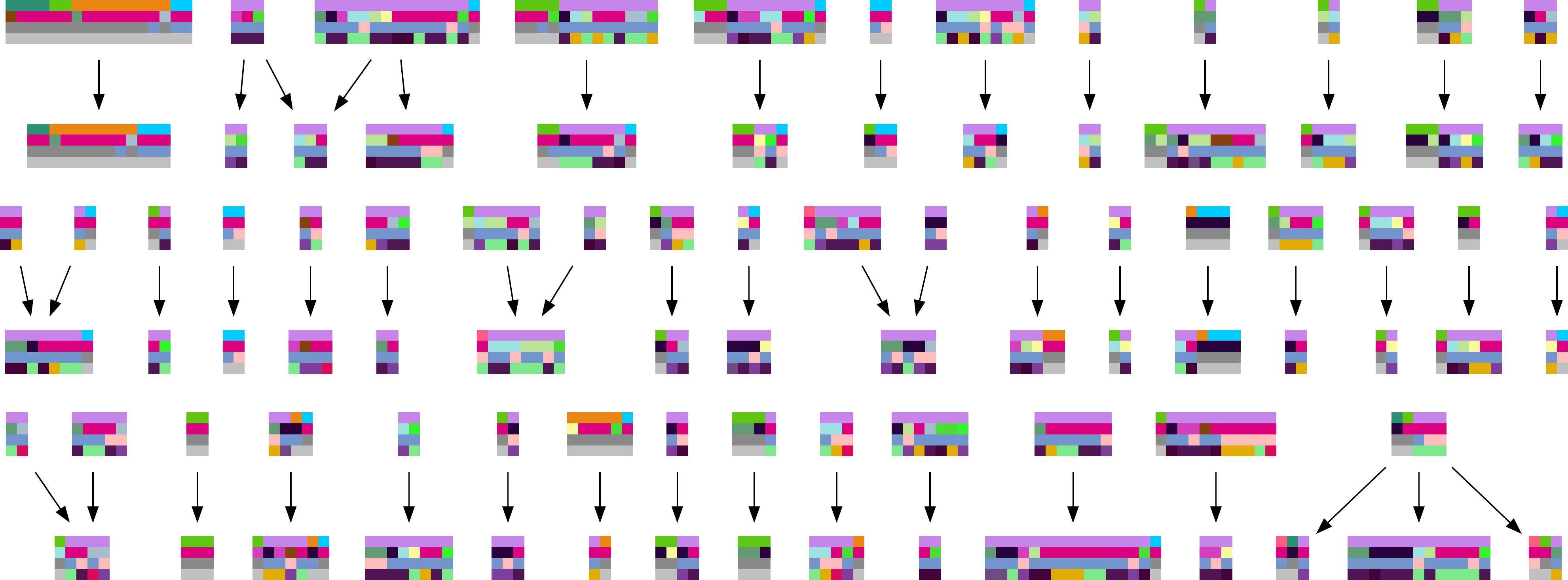}\par
      \caption{}
    \end{subfigure}
    \end{multicols} 
    
     \begin{multicols}{2}
    \begin{subfigure}{1\textwidth}
     \includegraphics[width=1\textwidth,height=\textheight,keepaspectratio]{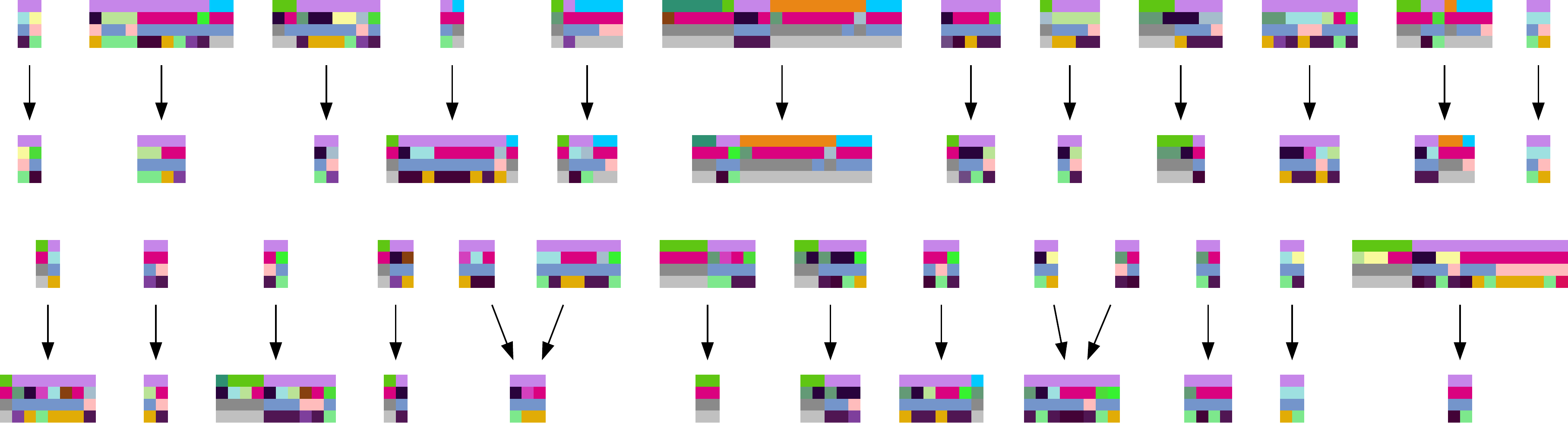}\par 
      \caption{}
    \end{subfigure}
    \end{multicols}
    
     \begin{multicols}{2}
    \begin{subfigure}{1\textwidth}
      \includegraphics[width=1\textwidth,height=\textheight,keepaspectratio]{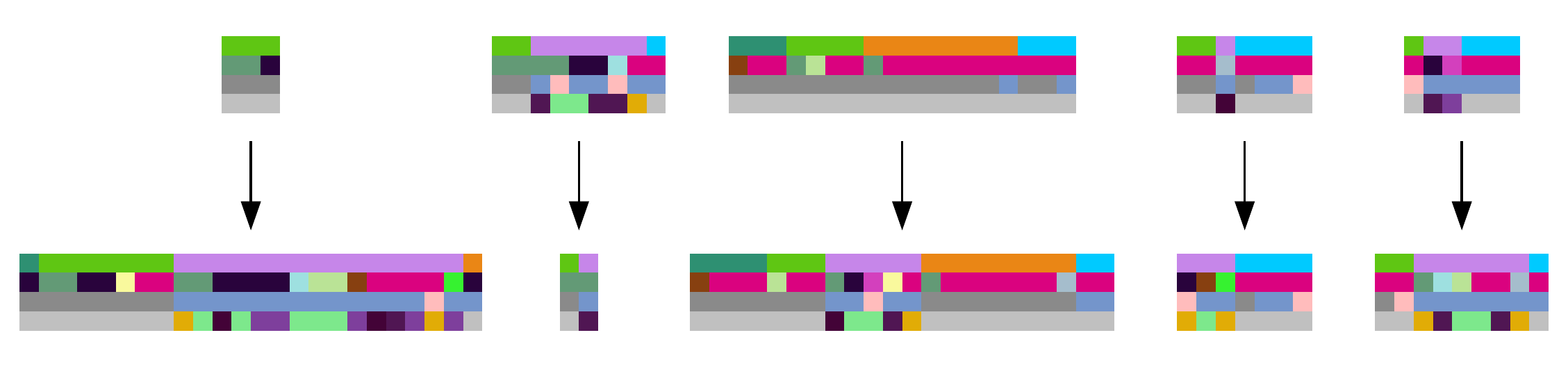}\par 
      \caption{}
    \end{subfigure}
    \end{multicols}
    
    \begin{multicols}{4}
            \begin{subfigure}[l]{0.27\textwidth}
      \includegraphics[width=1\textwidth,height=\textheight,keepaspectratio]{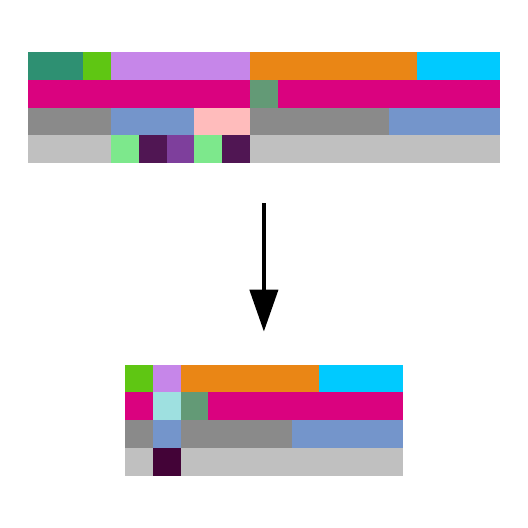}\par 
      \caption{}
    \end{subfigure}%
         \begin{subfigure}{0.26\textwidth}
      \includegraphics[width=.99\textwidth,height=\textheight,keepaspectratio]{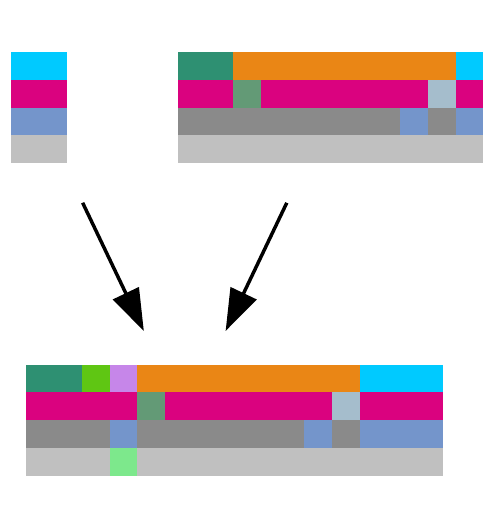}\par
      \caption{}
    \end{subfigure}%
    \begin{subfigure}{0.09\textwidth}
          \centering
      \includegraphics[width=0.95\textwidth,height=\textheight,keepaspectratio]{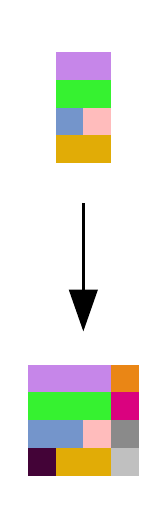}\par 
      \caption{}
    \end{subfigure}%
       \begin{subfigure}{0.43\textwidth}
      \includegraphics[width=0.99\textwidth,height=\textheight,keepaspectratio]{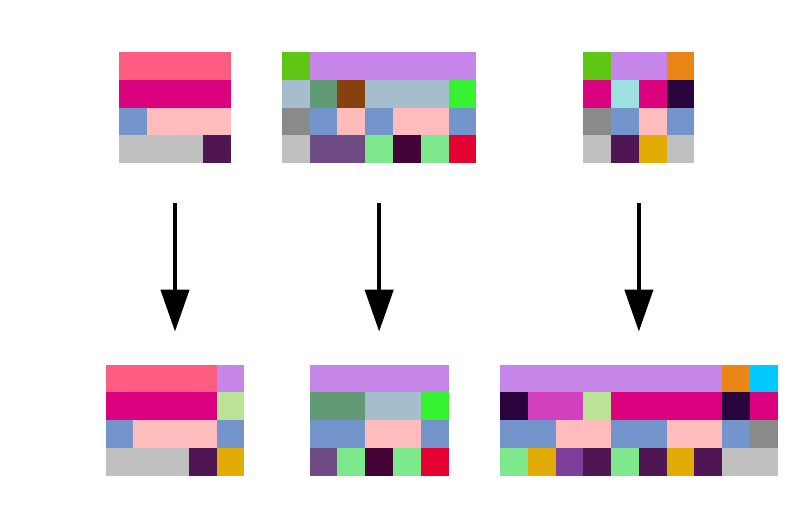}\par
      \caption{}
    \end{subfigure}
    \end{multicols}  
  \caption {
  {\bf Cluster evolution for FDR networks (continued from Fig. \ref{fig:netw_infomap_evol1}).}
  \label{fig:netw_infomap_evol2} (a) FI0009013403. (b) FI0009013429. (c) FI0009012843. (d) FI0009015309. (e) FI0009013312. (f) FI0009012413. (g) FI0009010391. 
}
\end{figure}

\begin{figure}[H]
    \centering
    \includegraphics[width=\textwidth]{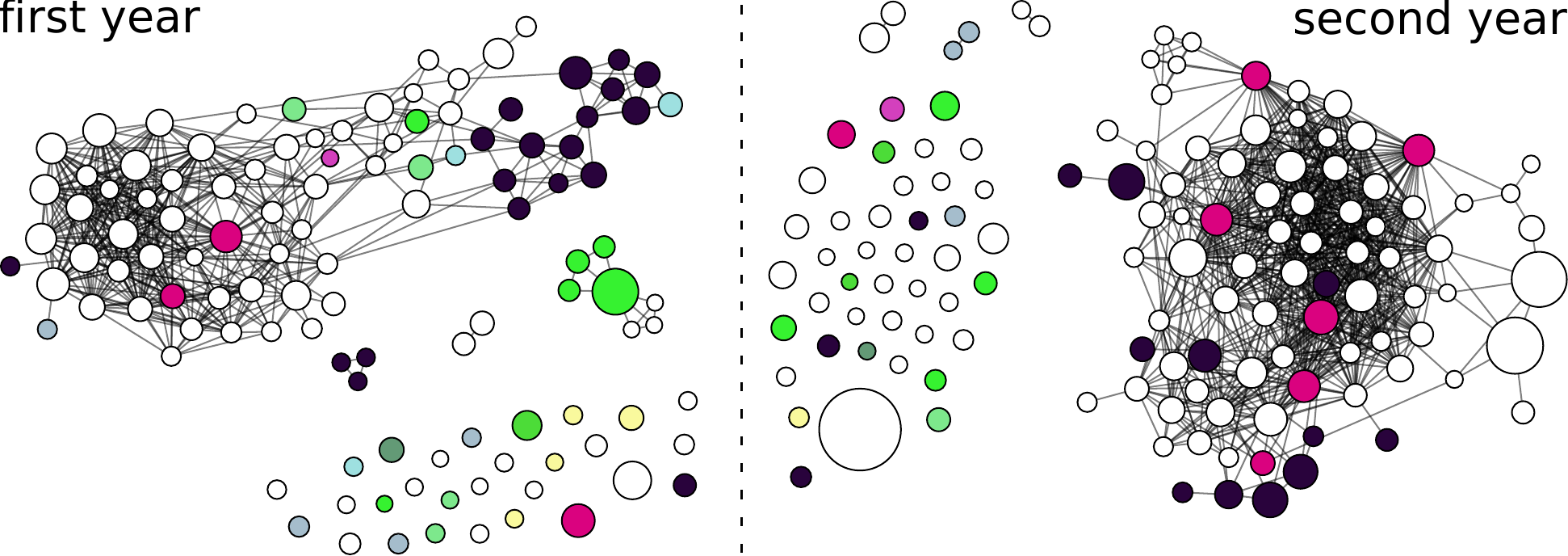}
    \caption{\textbf{Network of investor clusters with overexpressed attributes.} On the left-hand-side are the clusters observed in the first year after respective IPOs and right-hand-side, in the second year. Investor cluster nodes are connected with continuous links if they share statistically significant number of individual investors. 
    Node colours identify over-expressed geographic location within clusters.
\textbf{Geographic location:}
\protect\tikz{\protect\draw [black, thin, fill=cda027f] (4pt,4pt) circle (0.13);}  - Helsinki, 
\protect\tikz{\protect\draw [black, thin, fill=c29033c] (4pt,4pt) circle (0.13);} - South-West, 
\protect\tikz{\protect\draw [black, thin, fill=c639a76] (4pt,4pt) circle (0.13);} - Western-Tavastia, 
\protect\tikz{\protect\draw [black, thin, fill=c36f230] (4pt,4pt) circle (0.13);} - Central-Finland, 
\protect\tikz{\protect\draw [black, thin, fill=cf9f99d] (4pt,4pt) circle (0.13);} - Northern-Finland, 
\protect\tikz{\protect\draw [black, thin, fill=cbae396] (4pt,4pt) circle (0.13);} - Ostrobothnia,
\protect\tikz{\protect\draw [black, thin, fill=c9ee0e0] (4pt,4pt) circle (0.13) ;} - Rest-Uusimaa, 
\protect\tikz{\protect\draw [black, thin, fill=ca5bdcc] (4pt,4pt) circle (0.13);} - Eastern-Tavastia, 
\protect\tikz{\protect\draw [black, thin, fill=c4cdc38] (4pt,4pt) circle (0.13) ;} - Eastern-Finland, 
\protect\tikz{\protect\draw [black, thin, fill=cd340bd] (4pt,4pt) circle (0.13);} - South-East.
}
    \label{fig:postal_code_over_expressing_clusters}
\end{figure}

\begin{figure}[H]
\begin{center}
\includegraphics[width=0.6\linewidth]{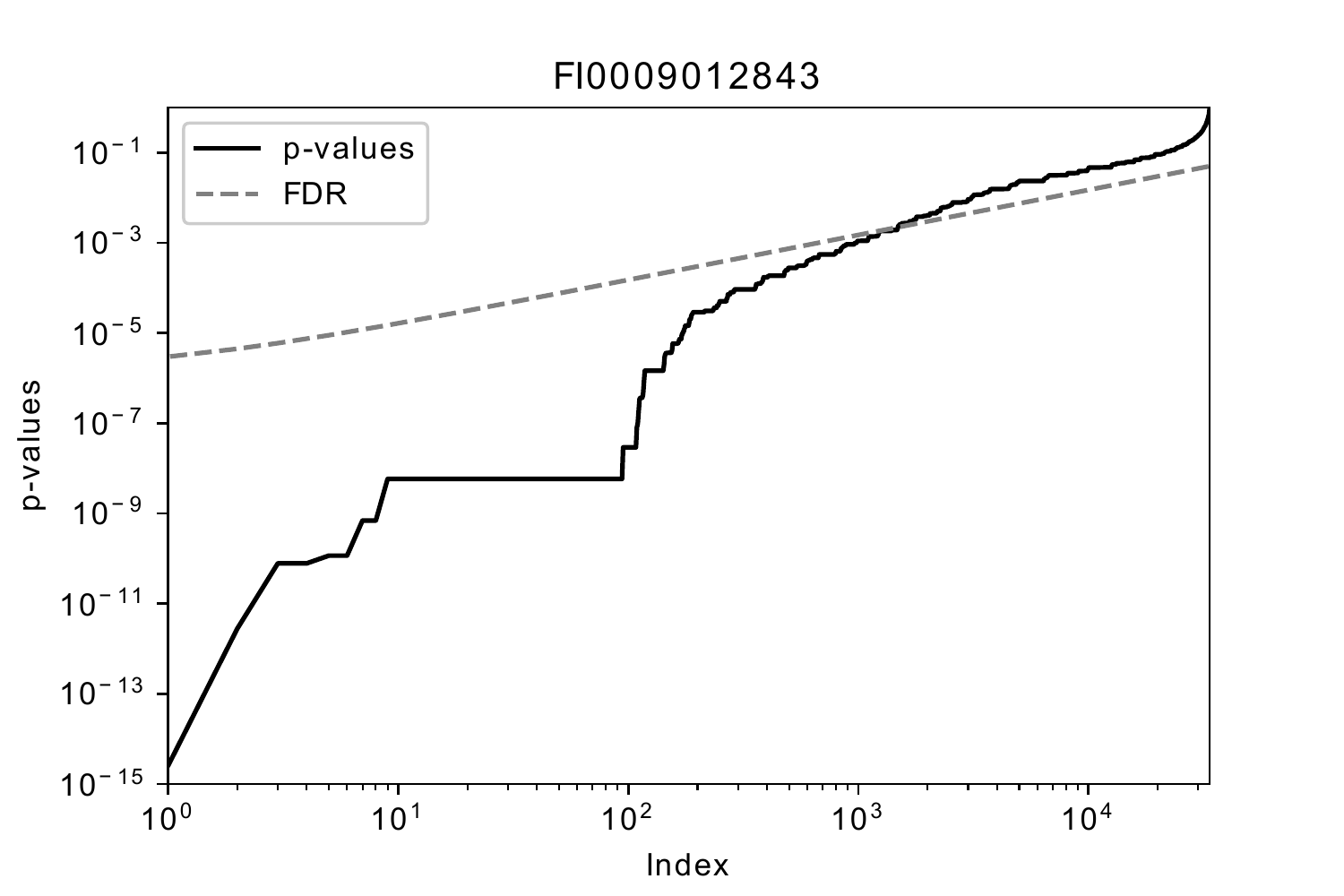}
\caption{{\bf Example of link validation for Kemira GrowHow (FI0009012843), first year after IPO, log-log scale.}
The number of observed synchronous trade co-occurrences: 33,595. 
The number of statistically validated links with the FDR correction: 1,481. 
}
\label{fig:link_validation}
\end{center}
\end{figure}

\title{Appendix to Clusters of investors around Initial Public Offering} 
\appendix
\maketitle
\setcounter{page}{1}
\counterwithin{figure}{section}
\counterwithin{table}{section}
The supplementary figures and tables are as follows:
\begin{itemize}
    \item Fig. \ref{fig:netw_infomap_evol1} and Fig. \ref{fig:netw_infomap_evol2} show the investor clusters that persisted from the first- into the second-year after IPO for eight securities. The clusters are shown as rectangle blocks that are composed of investors with four attributes: sector code, geographic location, gender and year of birth decade (for explanation see Fig. \ref{fig:com_expl} in the Article).
    Statistically significant overlapping cluster pairs are connected by the arrow from the first- to the second-year cluster.
    \item Fig. \ref{fig:overlaps_12} show the examples of security-wide investor cluster overlaps in the first (second) year after IPO. In the figure, each cluster has a statistically significant overlap with at least one cluster in a group, however, the arrows between the clusters are omitted for the simplification of the visualisation.
    \item Fig. \ref{fig:postal_code_over_expressing_clusters} shows the cluster groups with overexpressed geographical location attributes.
    \item  Tables \ref{tab_overexpression_grouped} and   \ref{tab_underexpression_grouped}  show the grouped clusters of overexpressed (underexpressed) attributes.
    \item Fig. \ref{fig:link_validation} illustrates the link validation procedure and shows the sorted $p$-values and the FDR correction threshold for Kemira GrowHow (FI0009012843). 
\end{itemize}

\section{The evolution of investor clusters on IPOs}
\label{app:ipo_clusters}

\begingroup
\begin{landscape}
\begin{figure}[!ht]
        \centering
            \captionsetup[subfigure]{justification=centering}
    \begin{multicols}{2}
         \begin{subfigure}{0.7\textwidth}
      \includegraphics[height=1.3\linewidth,keepaspectratio]{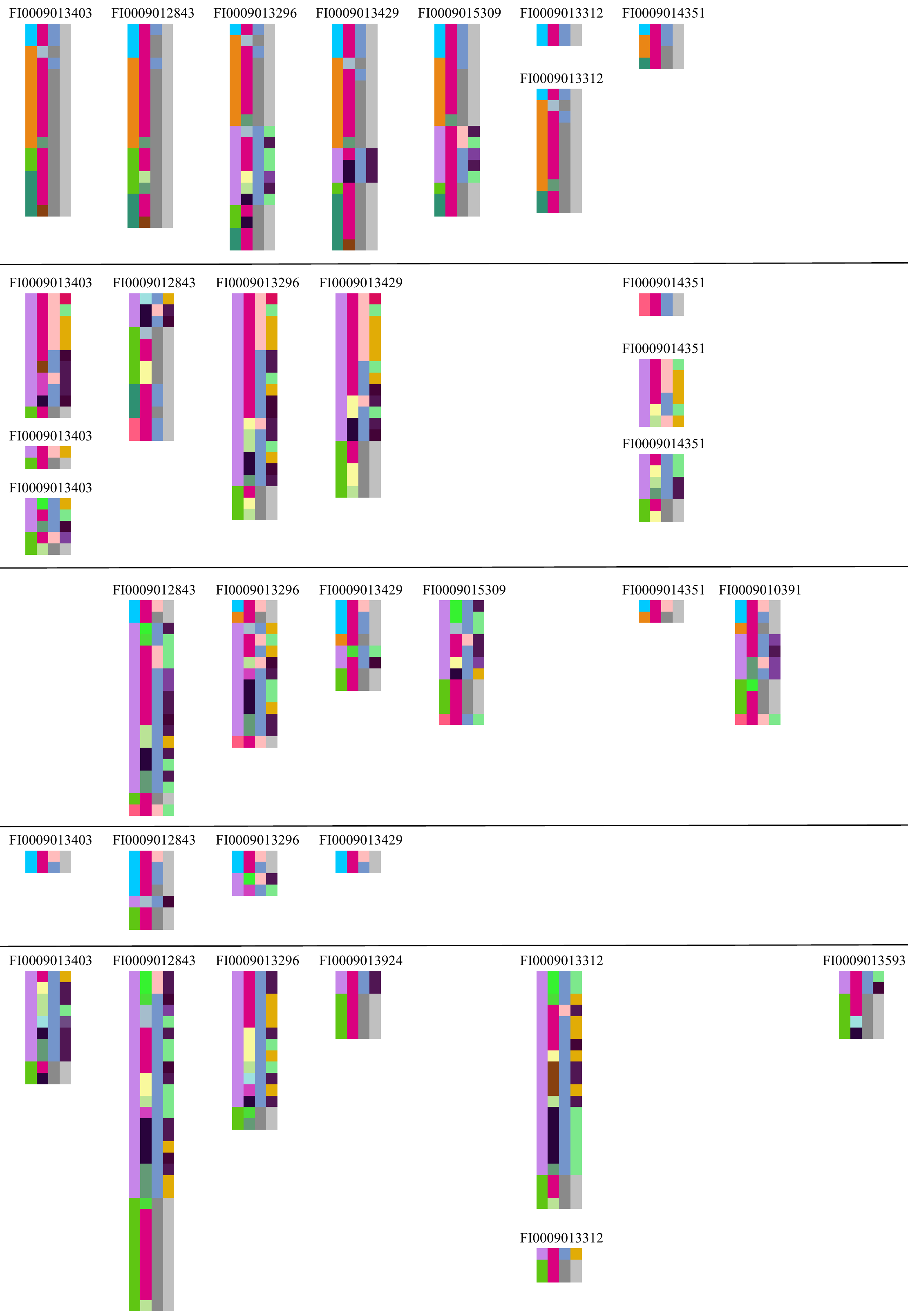}      \par
      \caption{\nth{1} year after IPO}
    \end{subfigure}%
    \begin{subfigure}{0.7\textwidth}
      \includegraphics[height=1.3\linewidth,keepaspectratio]{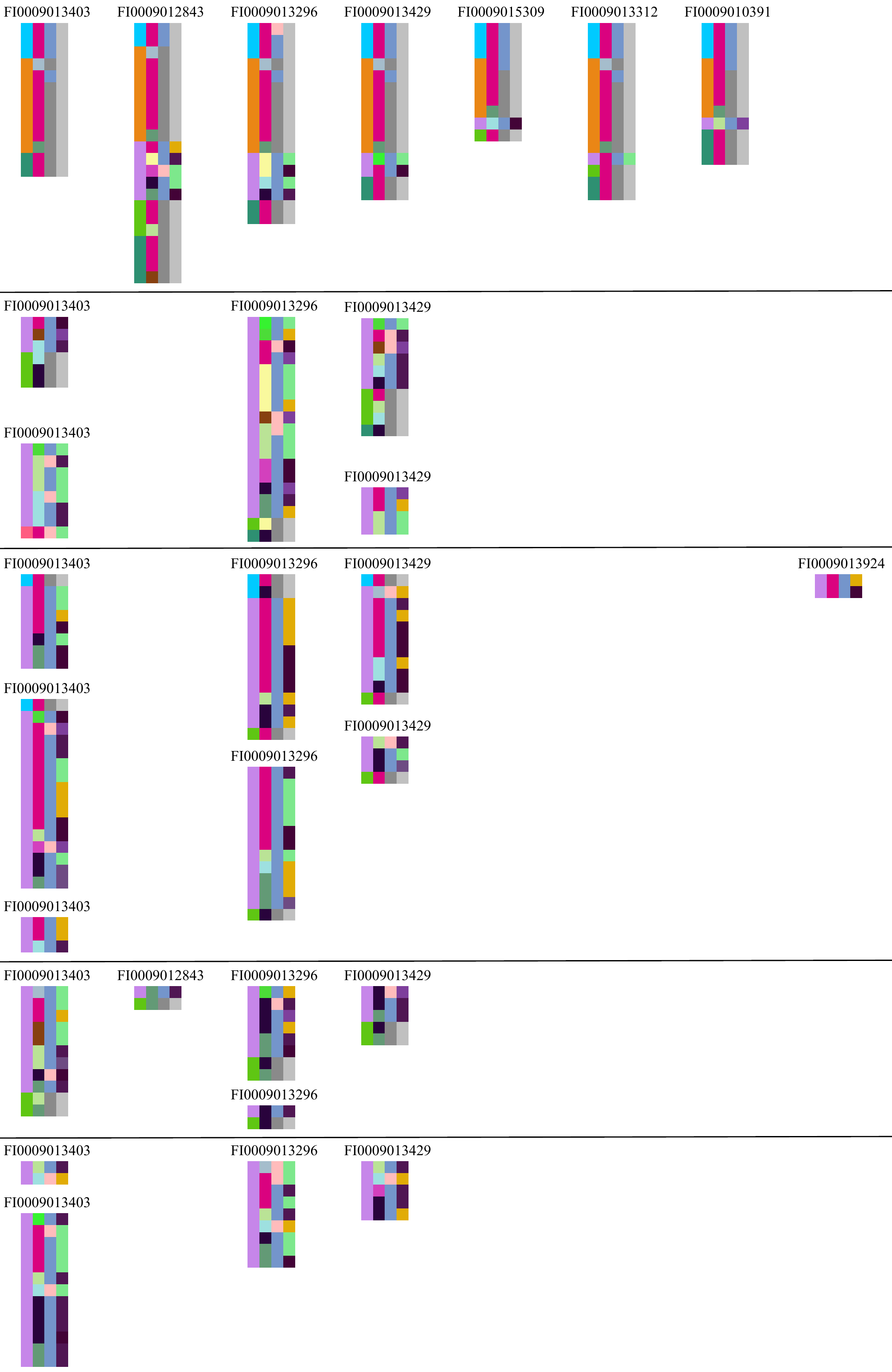} \par
      \caption{\nth{2} year after IPO}
    \end{subfigure}
    \end{multicols}
  \caption{
  {\bf Statistically significant cluster overlaps across multiple securities.}
  \label{fig:overlaps_12}  
(a) Overlapping clusters across multiple securities during the first year after an IPO. 
(b) Overlapping clusters across multiple securities during the second year after an IPO. 
A cluster is represented by the rectangle. 
Statistically overlapping cluster groups are separated by horizontal lines. 
Each cluster is composed of investors with four attributes: sector code, geographic location, gender and decade.
See the attribute colour mapping in Fig. \ref{fig:com_expl}.
}
\end{figure}  
\end{landscape}
\endgroup

\section{Over- and underexpressed attributes in groups of statistically similar clusters}

\LTcapwidth=\textwidth
\begin{longtabu} to \textwidth {Hc|Hrlr|HHrlr} 
\caption{\textbf{Overexpressed attributes in clusters' groups.} 
Here, the column 'Group' is a conventional name for a group of the statistically similar clusters with overexpressed attributes, where the left (right) side 'Y1' ('Y2') corresponds to the first (second) year after the IPO. 
'O-expr.' is the number of ISINs with overexpressed attributes in the same group, where the number of clusters  with overexpressed attributes is given in parentheses ().
'Attribute' is the name of the overexpressed attribute and '\# Attr.' is the number of ISINs in a group that overexpress this attribute, and the number of clusters that overexpress the attribute is given in parentheses ().
}\label{tab_overexpression_grouped}\\
\toprule
{} & \multirow{2}{*}{Group}  & \multicolumn{4}{c}{Y1}  & \multicolumn{5}{c}{Y2} \\
{} &                      & year\_x &                   O-expr. &           Attribute &   \# Attr. &  group\_y & year\_y &   O-expr. &          Attribute &   \# Attr. \\
\midrule
0  &   \multirow{15}{*}{1}    	 &     Y1 &  \multirow{15}{*}{24 (75)}   &    General-Government &  20 (32) &      1 &     Y2 &  \multirow{15}{*}{25 (83)} &   General-Government &  23 (42) \\
1  &      						 &     Y1 &					             &                No-Age &  16 (30) &      1 &     Y2 &					           &            No-Gender &  19 (39) \\
2  &      						 &     Y1 &					             &             No-Gender &  15 (29) &      1 &     Y2 &					           &               No-Age &  18 (39) \\
3  &      						 &     Y1 &					             &            South-West &  10 (15) &      1 &     Y2 &					           &  Financial-Insurance &  13 (22) \\
4  &      						 &     Y1 &					             &   Financial-Insurance &   8 (12) &      1 &     Y2 &					           &           Non-Profit &  13 (15) \\
5  &      						 &     Y1 &					             &            Non-Profit &    8 (9) &      1 &     Y2 &					           &           South-West &   8 (11) \\
6  &      						 &     Y1 &					             &          Ostrobothnia &    2 (2) &      1 &     Y2 &					           &             Helsinki &    5 (6) \\
7  &      						 &     Y1 &					             &              Helsinki &    2 (2) &      1 &     Y2 &					           &           Rest-World &    3 (3) \\
8  &      						 &     Y1 &					             &          Rest-Uusimaa &    2 (2) &      1 &     Y2 &					           &           Households &    2 (2) \\
9  &      						 &     Y1 &					             &       Central-Finland &    1 (1) &      1 &     Y2 &					           &        Non-Financial &    1 (1) \\
10 &      						 &     Y1 &					             &            South-East &    1 (1) &      1 &     Y2 &					           &                 1930 &    1 (1) \\
11 &      						 &     Y1 &					             &      Eastern-Tavastia &    1 (1) &      1 &     Y2 &					           &                 1940 &    1 (1) \\
12 &      						 &     Y1 &					             &         Non-Financial &    1 (1) &      1 &     Y2 &					           &                 1970 &    1 (1) \\
13 &      						 &     Y1 &					             &                Female &    1 (1) &        &        &                            &                      &          \\
14 &      						 &     Y1 &					             &                  1990 &    1 (1) &        &        &                            &                      &          \\
\hline
0  &         \multirow{2}{*}{2}  &     Y1 &    \multirow{2}{*}{6 (7)}    &       Central-Finland &    4 (4) &      2 &     Y2 &    \multirow{2}{*}{1 (1)}  &   \multirow{2}{*}{Central-Finland}     &     \multirow{2}{*}{1 (1)}  \\
1  &      						 &     Y1 &&            Non-Profit &    3 (3) &        &        &                            &                      &          \\
\hline
0  &                           3 &     Y1 &                        3 (3) &           South-West &    3 (3) &         &        &                            &                      &          \\
\hline
0  &       \multirow{2}{*}{4}    &     Y1 &   \multirow{2}{*}{2 (2)}    &  Financial-Insurance &    1 (1) &         &        &                            &                      &          \\
1  &                             &     Y1 &					             &           Households &    1 (1) &         &        &                            &                      &          \\
\hline
0  &        \multirow{3}{*}{5}   &        &                              &                      &          &      53 &     Y2 &    \multirow{3}{*}{2 (2)}  &  Financial-Insurance &    1 (1) \\
1  &      						 &        &                              &                      &          &      53 &     Y2 &                            &              No-Age  &    1 (1) \\
2  &      						 &        &                              &                      &          &      53 &     Y2 &                            &                 1990 &    1 (1) \\
\hline
0  &      					6	 &        &                              &                      &          &       0 &     Y2 &                      2 (2) &     Eastern-Tavastia &    2 (2) \\
\hline
0  &      					7    &        &                              &                      &          &      27 &     Y2 &                      2 (2) &  Financial-Insurance &    2 (2) \\
\hline

0  &      					8	 &     Y1 &                        1 (1) &           Rest-World &    1 (1) &      32 &     Y2 &                      1 (1) &           Rest-World &    1 (1) \\
\hline
0  &      				    9 	 &     Y1 &                        1 (1) &           South-West &    1 (1) &      13 &     Y2 &                      1 (1) &           South-West &    1 (1) \\
\hline
0  &      					10	 &     Y1 &                        1 (1) &                 1990 &    1 (1) &      17 &     Y2 &                      1 (1) &     Northern-Finland &    1 (1) \\

\hline
0  &      	\multirow{2}{*}{11}	 &     Y1 &   \multirow{2}{*}{1 (1)}     &           Households &    1 (1) &         &        &                            &                      &          \\
1  &      						 &     Y1 &                              &                 1980 &    1 (1) &         &        &                            &                      &          \\
\hline
0  &      	\multirow{2}{*}{12}	 &     Y1 &    \multirow{2}{*}{1 (1)}    &   General-Government &    1 (1) &         &        &                            &                      &          \\
1  &      						 &     Y1 &                              &               No-Age &    1 (1) &         &        &                            &                      &          \\
\hline
0  &         \multirow{2}{*}{13} &     Y1 &    \multirow{2}{*}{1 (1)}    &        Non-Financial &    1 (1) &         &        &                            &                      &          \\
1  &      						 &     Y1 &                              &            No-Gender &    1 (1) &         &        &                            &                      &          \\
\hline

0  &     14 &     Y1 &    1 (1) &  Financial-Insurance &    1 (1) &       &     &           &                   &       \\
\hline
0  &     15 &     Y1 &    1 (1) &           Rest-World &    1 (1) &       &     &           &                   &       \\
\hline
0  &     16 &     Y1 &    1 (1) &             Helsinki &    1 (1) &       &     &           &                   &       \\
\hline
0  &     17 &     Y1 &    1 (1) &     Northern-Finland &    1 (1) &       &     &           &                   &       \\
\hline
0  &     18 &     Y1 &    1 (1) &     Northern-Finland &    1 (1) &       &     &           &                   &       \\
\hline
0  &     19 &     Y1 &    1 (1) &     Northern-Finland &    1 (1) &       &     &           &                   &       \\
\hline
0  &     20 &     Y1 &    1 (1) &     Northern-Finland &    1 (1) &       &     &           &                   &       \\
\hline
0  &     21 &     Y1 &    1 (1) &      Eastern-Finland &    1 (1) &       &     &           &                   &       \\
\hline
0  &     22 &     Y1 &    1 (1) &      Central-Finland &    1 (1) &       &     &           &                   &       \\
\hline
0  &     23 &     Y1 &    1 (1) &     Eastern-Tavastia &    1 (1) &       &     &           &                   &       \\
\hline
0  &     24 &     Y1 &    1 (1) &     Eastern-Tavastia &    1 (1) &       &     &           &                   &       \\
\hline
0  &     25 &     Y1 &    1 (1) &     Eastern-Tavastia &    1 (1) &       &     &           &                   &       \\
\hline
0  &     26 &     Y1 &    1 (1) &     Western-Tavastia &    1 (1) &       &     &           &                   &       \\
\hline
0  &     27 &     Y1 &    1 (1) &         Ostrobothnia &    1 (1) &       &     &           &                   &       \\
\hline
0  &     28 &     Y1 &    1 (1) &         Ostrobothnia &    1 (1) &       &     &           &                   &       \\
\hline
0  &     29 &     Y1 &    1 (1) &         Rest-Uusimaa &    1 (1) &       &     &           &                   &       \\
\hline
0  &     30 &     Y1 &    1 (1) &                 1930 &    1 (1) &       &     &           &                   &       \\
\hline
0  &     31 &     Y1 &    1 (1) &                 1950 &    1 (1) &       &     &           &                   &       \\
\hline
0  &     32 &     Y1 &    1 (1) &                 1980 &    1 (1) &       &     &           &                   &       \\
\hline
0  &     33 &     Y1 &    1 (1) &                 1990 &    1 (1) &       &     &           &                   &       \\
\hline
0  &     34 &     Y1 &    1 (1) &                 1990 &    1 (1) &       &     &           &                   &       \\
\hline
0  &     35 &     Y1 &    1 (1) &                 1990 &    1 (1) &       &     &           &                   &       \\
\hline
0  &      	\multirow{2}{*}{36}	 &        &                              &                      &          &      43 &     Y2 &   \multirow{2}{*}{1 (1)}  &            No-Region &    1 (1) \\
1  &      						 &        &                              &                      &          &      43 &     Y2 &                            &           Rest-World &    1 (1) \\
\hline
0  &      \multirow{2}{*}{37}	 &        &                              &                      &          &      19 &     Y2 &    \multirow{2}{*}{1 (1)} &      Eastern-Finland &    1 (1) \\
1  &      						 &        &                              &                      &          &      19 &     Y2 &                            &               Female &    1 (1) \\
\hline
0  &      \multirow{2}{*}{38}    &        &                              &                      &          &      47 &     Y2 &    \multirow{2}{*}{1 (1)} &                 Male &    1 (1) \\
1  &      						 &        &                              &                      &          &      47 &     Y2 &                            &                 1940 &    1 (1) \\
\hline
0  &      \multirow{2}{*}{39}    &        &                              &                      &          &      69 &     Y2 &    \multirow{2}{*}{1 (1)} &     Western-Tavastia &    1 (1) \\
1  &      						 &        &                              &                      &          &      69 &     Y2 &                            &     1970 &    1 (1) \\
\hline
0  &         \multirow{2}{*}{40} &        &                              &                      &          &      68 &     Y2 &    \multirow{2}{*}{1 (1)} &            No-Region &    1 (1) \\
1  &      						 &        &                              &                      &          &      68 &     Y2 &                            &           Rest-World &    1 (1) \\
\hline
0  &     41  &     &           &                   &       &     25 &     Y2 &    1 (1) &           Households &    1 (1) \\
\hline
0  &    42   &     &           &                   &       &      9 &     Y2 &    1 (1) &           Households &    1 (1) \\
\hline
0  &      43 &     &           &                   &       &     39 &     Y2 &    1 (1) &           Households &    1 (1) \\
\hline
0  &    44   &     &           &                   &       &     14 &     Y2 &    1 (1) &           Non-Profit &    1 (1) \\
\hline
0  &     45  &     &           &                   &       &     67 &     Y2 &    1 (1) &  Financial-Insurance &    1 (1) \\
\hline
0  &     46  &     &           &                   &       &     61 &     Y2 &    1 (1) &           Rest-World &    1 (1) \\
\hline
0  &     47  &     &           &                   &       &     24 &     Y2 &    1 (1) &             Helsinki &    1 (1) \\
\hline
0  &    48   &     &           &                   &       &     54 &     Y2 &    1 (1) &           South-West &    1 (1) \\
\hline
0  &     49  &     &           &                   &       &     64 &     Y2 &    1 (1) &           South-West &    1 (1) \\
\hline
0  &     50  &     &           &                   &       &      4 &     Y2 &    1 (1) &           South-East &    1 (1) \\
\hline
0  &     51  &     &           &                   &       &     48 &     Y2 &    1 (1) &         Ostrobothnia &    1 (1) \\
\hline
0  &      52 &     &           &                   &       &     46 &     Y2 &    1 (1) &      Central-Finland &    1 (1) \\
\hline
0  &     53  &     &           &                   &       &     63 &     Y2 &    1 (1) &      Central-Finland &    1 (1) \\
\hline
0  &     54  &     &           &                   &       &     41 &     Y2 &    1 (1) &      Central-Finland &    1 (1) \\
\hline
0  &     55  &     &           &                   &       &     40 &     Y2 &    1 (1) &      Eastern-Finland &    1 (1) \\
\hline
0  &     56  &     &           &                   &       &     49 &     Y2 &    1 (1) &     Eastern-Tavastia &    1 (1) \\
\hline
0  &     57  &     &           &                   &       &     35 &     Y2 &    1 (1) &            No-Region &    1 (1) \\
\hline
0  &     58  &     &           &                   &       &     34 &     Y2 &    1 (1) &                 1910 &    1 (1) \\
\hline
0  &     59  &     &           &                   &       &     70 &     Y2 &    1 (1) &                 1920 &    1 (1) \\
\hline
0  &    60   &     &           &                   &       &     23 &     Y2 &    1 (1) &                 1970 &    1 (1) \\
\hline
0  &    61   &     &           &                   &       &      3 &     Y2 &    1 (1) &                 1970 &    1 (1) \\
\hline
0  &     62  &     &           &                   &       &     52 &     Y2 &    1 (1) &                 1980 &    1 (1) \\
\hline
0  &    63   &     &           &                   &       &      8 &     Y2 &    1 (1) &                 1980 &    1 (1) \\
\hline
0  &     64  &     &           &                   &       &     21 &     Y2 &    1 (1) &                 1980 &    1 (1) \\
\hline
0  &     65  &     &           &                   &       &     71 &     Y2 &    1 (1) &                 1980 &    1 (1) \\
\hline
0  &     66  &     &           &                   &       &     30 &     Y2 &    1 (1) &                 1990 &    1 (1) \\
\hline
0  &    67   &     &           &                   &       &     29 &     Y2 &    1 (1) &                 1990 &    1 (1) \\
\hline
0  &    68   &     &           &                   &       &     59 &     Y2 &    1 (1) &                 1990 &    1 (1) \\
\hline
0  &     69  &     &           &                   &       &      7 &     Y2 &    1 (1) &                 1990 &    1 (1) \\
\hline
0  &   70    &     &           &                   &       &     37 &     Y2 &    1 (1) &                 1990 &    1 (1) \\
\hline
0  &     71  &     &           &                   &       &     42 &     Y2 &    1 (1) &                 2000 &    1 (1) \\
\hline
0  &     72  &     &           &                   &       &     55 &     Y2 &    1 (1) &                 2000 &    1 (1) \\
\bottomrule
\end{longtabu}

\newpage
\begin{longtabu} to \textwidth {Hc|Hllr|HHllr}
\caption{\textbf{Underexpressed attributes in clusters' groups.}
Here, the column 'Group' is a conventional name for a group of the statistically similar clusters with underexpressed attributes, where the left (right) side 'Y1' ('Y2') corresponds to the first (second) year after the IPO. 
'U-expr.' is the number of ISINs with underexpressed attributes in the same group, where the number of clusters  with underexpressed attributes is given in parentheses ().
'Attribute' is the name of the underexpressed attribute and '\# Attr.' is the number of ISINs in a group that underexpress this attribute, and the number of clusters that underexpress the attribute is given in parentheses ().
}\label{tab_underexpression_grouped}\\
\toprule
{} & \multirow{2}{*}{Group}  & \multicolumn{4}{c}{Y1}  & \multicolumn{5}{c}{Y2} \\
{} &                      & year\_x &                   U-expr. &           Attribute &   \# Attr. &  group\_y & year\_y &   U-expr. &          Attribute &   \# Attr. \\
\midrule
0 &     \multirow{7}{*}{1} &     Y1 &   \multirow{2}{*}{14 (22)}   &    Households    &  11 (19) &      1 &     Y2 & \multirow{7}{*}{ 19 (41)} &           Households &  17 (34) \\
1 &                        &     Y1 &                             &            Male   &  11 (14) &      1 &     Y2 &                           &                 Male &  14 (27) \\
2 &                        &        &                             &                &          &      1 &     Y2 &                           &  Financial-Insurance &    1 (1) \\
3 &                        &        &                             &                &          &      1 &     Y2 &                           &             Helsinki &    1 (1) \\
4 &                        &        &                             &                &          &      1 &     Y2 &                           &                1940  &    1 (1) \\
5 &                        &        &                             &                &          &      1 &     Y2 &                           &                 1950 &    1 (1) \\
6 &                        &        &                             &                &          &      1 &     Y2 &                           &               No-Age &    1 (1) \\
\hline
0 &                      2 &     Y1 &                       2 (2) &       Helsinki &    2 (2) &        &        &                           &                   &       \\
\hline 
0 &                      3  &        &                            &                &           &      4 &     Y2 &                     2 (2) &             Helsinki &    2 (2) \\
\hline
0 &      4 &     Y1 &    1 (1) &  Households &    1 (1) &       &     &           &                   &       \\
\hline
0 &      5 &     Y1 &    1 (1) &      No-Age &    1 (1) &       &     &           &                   &       \\
\hline
0 &      6 &     Y1 &    1 (1) &    Helsinki &    1 (1) &       &     &           &                   &       \\
\hline
0 &     \multirow{2}{*}{7} &     Y1 &    \multirow{2}{*}{1 (1)} &    Helsinki &    1 (1) &       &     &           &                   &       \\
1 &                        &     Y1 &   &  South-West &    1 (1) &       &     &           &                   &       \\
\hline
0 &     \multirow{2}{*}{8} &     &           &          &       &      2 &     Y2 &    \multirow{2}{*}{1 (1)} &               No-Age &    1 (1) \\
1 &                        &     &           &          &       &      2 &     Y2 &    &            No-Gender &    1 (1) \\

\bottomrule
\end{longtabu}

\section{Link validation with the FDR correction}
\label{app:link_validation}

\newpage


\end{document}